%% file: main.tex
\documentclass[11pt]{article}

\usepackage[T1]{fontenc}
\usepackage[utf8]{inputenc}
\usepackage{lmodern}
\usepackage[a4paper,margin=1in]{geometry}
\usepackage{microtype}
\usepackage{amsmath,amssymb}
\usepackage{graphicx}
\usepackage{placeins}
\usepackage{booktabs}
\usepackage{array}
\usepackage{xcolor}
\usepackage{caption}
\usepackage{subcaption}
\usepackage{enumitem}
\usepackage{natbib}
\usepackage{url}
\usepackage{hyperref}
\usepackage{cleveref}

\hypersetup{
  colorlinks=true,
  linkcolor=blue!45!black,
  citecolor=blue!45!black,
  urlcolor=blue!55!black,
  pdftitle={Latent Molecular Rollout Towards Causal World Models},
  pdfauthor={Wang Rui}
}
\newcommand{\pka}{\mathrm{p}K_{\mathrm a}}
\newcommand{\MAE}{\mathrm{MAE}}

\newcommand{\cmark}{\textcolor{green!45!black}{Pass}}
\newcommand{\xmark}{\textcolor{red!55!black}{Fail}}

\title{Latent Molecular Rollout Towards Causal World Models\\[0.4em]\large Direction is Learnable, Amplitude Control is Readout-Dependent}
\author{
  Wang Rui$^{*}$\\[0.3em]
  Department of Chemical Engineering, Tsinghua University\\[0.3em]
  $^{*}$Corresponding author: \href{mailto:r-wang18@tsinghua.org.cn}{r-wang18@tsinghua.org.cn}\\
  ORCID: \href{https://orcid.org/0000-0002-2362-3214}{0000-0002-2362-3214}
}
\date{November 14, 2026}

\begin{document}
\maketitle

\begin{abstract}
A molecular world model should predict how a structural change moves a property, not merely label a static molecule. We build a latent molecular rollout on an 8B Qwen3 backbone trained on 9{,}202 aqueous $\pka$ pairs, and ask how far such a rollout reaches toward causal control by auditing three capabilities that are usually conflated.

Removing textual shortcuts, latent inference reaches test MAE $2.087$, $0.276$ below direct molecular readout (95\%~CI $[-0.352,-0.203]$). A target-conditioned policy learns displacement direction (median cosine $0.577$; $0.547$ on a pre-registered 151-pair expansion) but not its magnitude. That boundary is geometric, not a fitting failure: nine independently fitted readouts stay two to three orders of magnitude below the random-direction scale and none recovers amplitude, whereas the pathway-co-trained head does. Neither input-slot editing nor final-state injection changes any of 80 decoded structures.

The rollout thus supplies a learnable displacement direction and locates, rather than merely reports, what it lacks: amplitude is set by readout geometry, and decoded transfer by the latent-to-structure interface.
\end{abstract}

\section{Introduction}
\label{sec:introduction}

A causal molecular world model must support more than accurate static property prediction: it should enable targeted interventions, predict how structural changes alter physical properties, and generate rollout trajectories toward desired states \citep{Ha2018WorldModels,Scholkopf2021Causal}. Latent-space molecular models have made rapid progress toward this goal, with smooth interpolations and strong scalar prediction performance often taken as evidence of controllable representations \citep{Eckmann2022LIMO,Soares2025SMITED,Hao2025Coconut}.

Yet causal controllability cannot be reduced to a single scalar metric. Progress toward a deployable molecular world model requires three separable capabilities to hold simultaneously, each of which must be verified independently rather than assumed. First, predictive accuracy must arise from the intended molecular representation pathway, not from textual shortcuts in the input. Second, latent displacements must support both reliable directional alignment and calibrated property amplitude, rather than being validated by cosine similarity alone. Third, control demonstrated in latent space must transfer reliably to decoded structural outputs, rather than remaining confined to hidden-state manipulations. Conflating these tiers leads to overstated claims of causal control: a model can excel at one tier while failing entirely at the next. Causal representation learning and activation-steering work have separately noted that smoothness and controllability are distinct axes \citep{Scholkopf2021Causal,Park2024Linear,Turner2024ActAdd}, and probing studies warn that a successful readout need not imply use of the recovered information in downstream computation \citep{Belinkov2022Probing}. We therefore structure our evaluation as a three-tier audit. The protocol is deliberately model-agnostic: each tier is an independent test that any latent molecular model must pass before its interventions can be called controllable, and the diagnostics we introduce---shortcut removal, direction/amplitude separation with a readout-family control, and decoded-transfer checks---apply to other architectures. We apply it here to one system to delineate what it can and cannot support.

In this work, we build a latent molecular rollout system on an 8B Qwen3 backbone, using aqueous experimental $\pka$ as a physically measurable anchor. To rigorously audit how far it advances toward causal control, we evaluate three separable capabilities independently---a three-tier audit. \textbf{Tier 1 (scalar prediction):} does the repaired latent-inference pathway carry predictive value through the intended molecular channel, or does accuracy come from textual shortcuts? \textbf{Tier 2 (direction vs.\ amplitude):} can a target-conditioned policy learn displacement direction, and does that direction carry a calibrated property amplitude under different readouts? \textbf{Tier 3 (decoded control):} does latent-state intervention transfer to decoded molecular structure? Each tier answers a distinct question about causal capability; a pass at one tier does not establish the next.

Autonomous and tool-augmented chemical systems offer a complementary validation model. Coscientist and ChemCrow connect language models to chemistry tools \citep{Boiko2023Coscientist,Bran2024ChemCrow}, while A-Lab integrates computational planning with robotic synthesis \citep{Szymanski2023ALab}. In these systems, confidence in a proposal is anchored externally by computation, execution, or measurement. We study the complementary question of internal, pre-validation auditability: whether latent geometry exposes a meaningful direction and a controllable magnitude before a proposal reaches an external oracle. External validation loops ask whether candidates can be validated at scale; internal audits ask whether the proposed move can be inspected beforehand. Neither route replaces the other. We do not run an external validation loop here. Our question is whether a latent displacement is inspectable before any oracle is called, not whether an external oracle can filter proposals; a head-to-head comparison would require a shared candidate budget and an oracle-cost model, which we do not attempt.

We instantiate this program using aqueous $\pka$ as a physically grounded anchor. Unlike fragment-additive descriptors such as logP or QED, which can be well approximated by counting substructural motifs, $\pka$ arises from the molecule's global electronic structure, modulated by inductive, resonance, and solvation effects. This makes it a stricter test of causal representation: a model cannot reliably guide property-altering structural changes by memorizing fragment correlations alone; it must capture how electronic reorganization shifts the dissociation equilibrium.

For the conventional dissociation equilibrium $\mathrm{HA}\rightleftharpoons\mathrm{H}^{+}+\mathrm{A}^{-}$, the acidity constant and its logarithmic form are
\begin{equation}
 K_{\mathrm a}=\frac{a_{\mathrm H^+}a_{\mathrm A^-}}{a_{\mathrm{HA}}},
 \qquad \pka=-\log_{10}K_{\mathrm a},
\end{equation}
where $a$ denotes activity \citep{IUPACAcidityConstant}. Under the same solvent and experimental conditions, a lower $\pka$ corresponds to a stronger acid. In the ideal dilute approximation, the Henderson--Hasselbalch relation, $\mathrm{pH}=\pka+\log_{10}([\mathrm A^-]/[\mathrm{HA}])$, connects this quantity to the balance between protonated and deprotonated forms. Thus $\pka$ supplies a chemically interpretable intervention target rather than an arbitrary regression label. It is not a solvent-independent molecular constant: solvent and experimental conditions matter, so our in-domain and out-of-domain evaluations must be interpreted separately \citep{Yang2020HolisticPKa}. The iBonD resource supplies the aqueous-phase experimental measurements used here \citep{YangIbonD,Wang2026iBonD}.

SMILES (Simplified Molecular Input Line Entry System) encodes a molecular graph as a character string \citep{Weininger1988SMILES}. This makes it a convenient interface for molecular language models, but also creates an information-shortcut risk: when the molecule is included verbatim in the textual query, a predictor can use the answer-bearing text instead of the intended molecular representation. We therefore distinguish text tokens from molecular features throughout, remove query SMILES from the repaired regression prompt, and test the resulting pathway with explicit controls.

\begin{figure}[t]
\centering
\includegraphics[width=0.92\textwidth]{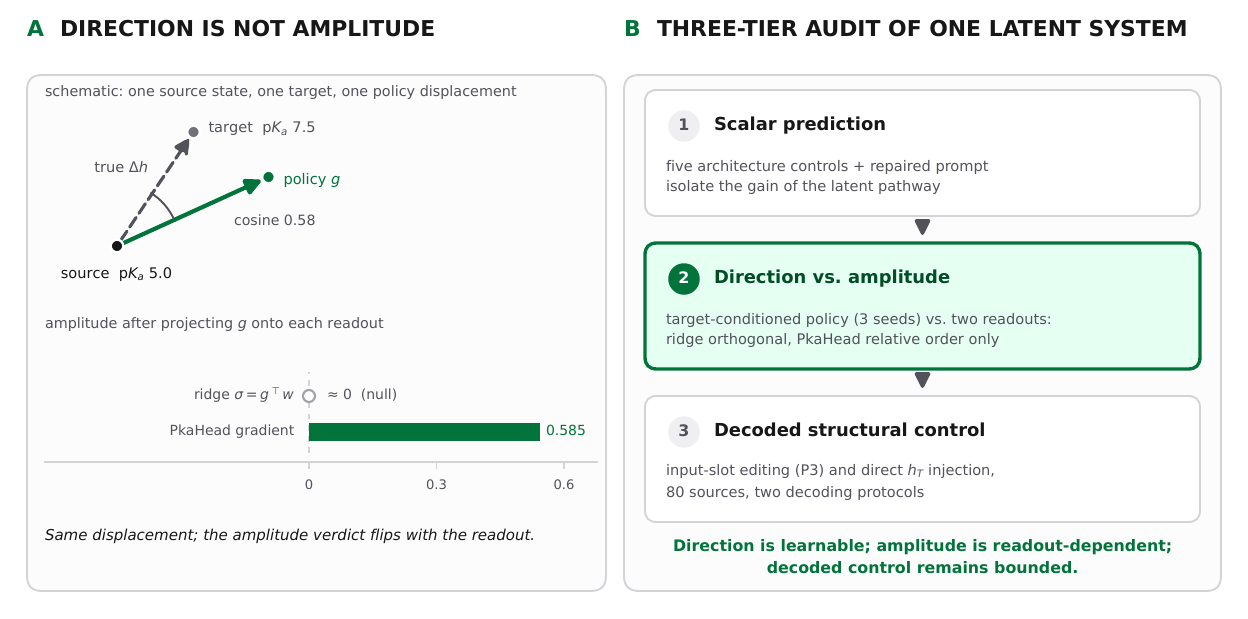}
\caption{Concept and audit overview. Left: schematic of one source state, its true displacement toward the target, and the learned policy displacement (median cosine $0.58$; angles not to scale); the same displacement projects to near-zero amplitude under the frozen ridge readout and to relative amplitude order under the \texttt{PkaHead} gradient. Right: the paper therefore audits scalar prediction, latent direction/amplitude, and decoded control as separate questions; a strong value at one stage does not establish the next. Generation is a downstream application of decoded control, not a fourth independent audit.}
\label{fig:intro_overview}
\end{figure}

We build on LatentChem, a molecular language architecture with a continuous latent-inference pathway and a scalar \texttt{PkaHead} \citep{Ye2026LatentChem}, to construct an intervention-responsive rollout and a target-conditioned generation path. Five architecture variants are compared using the same held-out $\pka$ criterion. The central question is deliberately narrower than ``is this a world model?'': \emph{can the same latent policy learn both the direction and the amplitude of a target-conditioned molecular change?} \Cref{fig:intro_overview} summarizes the system and the three audits that keep scalar prediction, latent geometry, and decoded control conceptually separate.

Our evaluation combines scalar controls, interpolation and multi-step accumulation, direct interventions, a target-conditioned direction policy, and generation. The policy head is trained on training-split molecular pairs and evaluated on 43 held-out pairs with separate prespecified direction and amplitude criteria. Fixed-step sensitivity tests whether the ridge null is merely a scale-choice artifact, a nonlinear \texttt{PkaHead}-gradient reread tests whether it is a readout artifact, and an explicit amplitude loss tests whether adding ridge-targeted supervision restores joint control. Naive gradient alignment, a frozen ridge readout, readout-free PCA, and exploratory shared-direction projections remain supporting diagnostics. The generation path uses paired single-sample comparisons, a best-constant predictor, a no-op reference, and Best-of-8 oracle selection to quantify realized improvement and remaining headroom.

The contributions are fourfold.
\begin{enumerate}[leftmargin=1.5em]
  \item \textbf{Build a latent rollout that supports direction and accumulation.} Continuous latent inference attains $2.087$ test MAE, with a $0.256$ residual improvement after random-position control and a paired C-versus-B$'$ bootstrap interval excluding zero. This residual establishes predictive value over random positions, not molecule-specific latent trajectories or a unique shared-query mechanism. G1 is monotonic for 38/38 pairs, G2 reduces median absolute target error from $2.000$ to $0.251$, G5 reaches $0.676$ sign accuracy, and the target-conditioned policy attains seed-median cosines of $0.469$--$0.539$ (median $0.482$).
  \item \textbf{Show that amplitude control is readout-dependent, not absent.} On the same 43 held-out pairs, the policy ensemble reaches cosine $0.577$ and $27/43=0.628$ pairs exceed $0.3$ (exact 95\% CI $[0.467,0.770]$). The amplitude null previously obtained with the selected frozen ridge is mechanically induced: that readout's median cosine to true displacements is $1.18\times10^{-5}$, so an amplitude correlation near zero is expected regardless of the policy. The nonlinear \texttt{PkaHead} gradient gives correlation $0.585$ and satisfies the prespecified joint criterion; the correlation draws on both step length ($0.355$) and unit direction ($0.460$), and the scale is compressed rather than absent (median $2.33$ against a true $3.03$), with no reduction in the held-out residual from a global factor fitted on training pairs. Relative amplitude ordering is therefore recovered under a nonlinear readout while absolute calibration is not established. Across nine independently fitted readouts on the same latents---ridge penalties, neural heads, a label-shuffled probe, and a half-data probe---gradient--displacement alignment stays at $10^{-5}$ and no readout recovers amplitude, locating the dependence in readout geometry rather than in one head.
  \item \textbf{Audit two latent control surfaces.} In an 80-source audit, optimizing the  input slot embeddings reaches median target-state cosine $1.0000$ and relative error $0.0081$, yet greedy and sampled decoding change only $1/80$ and $0/80$ sequences. Direct injection into the final \texttt{<end\_latent>} hidden state, without slot optimization, changes $0/80$ sequences for every combination of four steps and two decoding protocols; median relative displacements span $0.0675\%$--$1.617\%$. The initial slot blocks are nearly molecule-invariant, but this describes only their pre-forward initialization. Together, the audits bound the tested control surfaces and magnitudes rather than establishing a decoder-wide bottleneck; ChemToken editing remains untested.
  \item \textbf{Bound generation improvement explicitly.} Generation is the downstream application of decoded control rather than a fourth independent audit. Oracle-best SFT improves paired direction accuracy and MAE over gen2-final, but does not establish a direction-accuracy advantage over the best constant on the same rows, and a fixed three-epoch budget ablation does not change that conclusion. Offline DPO remains below the $0.72$ direction design reference at both two epochs ($0.7137$) and four epochs ($0.6353$) and does not detect an advantage over the train-selected constant ($p=0.2215$ and $p=0.3742$); its $0.0039$ copy rate is within the $0.15$ design bound. Best-of-8 reaches $0.800$ direction accuracy and $1.313$ MAE under oracle selection, quantifying headroom rather than deployable performance.
\end{enumerate}

\section{Related work}
\label{sec:related}

World models and causal representation learning motivate evaluating latent states by prediction, intervention, and rollout rather than by representation probes alone \citep{Ha2018WorldModels,Hafner2025Dreamer,Scholkopf2021Causal}. In chemistry, continuous latent reasoning and molecular encoder--decoder models provide editable representations and latent-space design \citep{Hao2025Coconut,Ye2026LatentChem,Soares2025SMITED,Eckmann2022LIMO}, while linear representation and activation-steering work suggests that directions can support controlled interventions \citep{Park2024Linear,Turner2024ActAdd}. These ideas motivate our question: whether a learned latent state supports directional and calibrated $\pka$ control.

\paragraph{External validation loops.} A complementary line couples proposal generation to computation or execution. Coscientist and ChemCrow route language-model proposals through chemistry tools \citep{Boiko2023Coscientist,Bran2024ChemCrow}, A-Lab closes the loop with robotic synthesis \citep{Szymanski2023ALab}, and related systems use DFT or high-throughput calculation as the acceptance oracle. Many of these systems integrate retrosynthesis planners to validate synthetic accessibility, which answers a distinct question: given a fixed target structure, how to synthesize it from available precursors. In all of these, a proposal's credibility is established after an external call. Our work addresses the upstream, property-directed design problem: how a starting molecule should be structurally altered to achieve a desired physical property, before any synthesis planning is invoked. The two lines are complementary in that an internal audit of latent property rollout could in principle feed candidate structures into a retrosynthesis pipeline for full loop closure; we do not test that integration here.

Experimental $\pka$ supervision comes from iBonD and related $\pka$-prediction resources \citep{YangIbonD,Wang2026iBonD,Yang2020HolisticPKa,Pan2021MolGpKa,Luo2024UniPKa,Wang2026PkaAugmentation}. Our $2.087$ test MAE is an internal architectural result on a fixed split, not a same-split comparison with dedicated predictors. We use $\pka$ as a measurable anchor for intervention response and directional control.

Probing studies warn that a successful readout need not imply use of the recovered information in downstream computation \citep{Belinkov2022Probing}. We therefore report scalar accuracy, direction, amplitude, and decoded control separately. Binary outcomes use exact McNemar tests and paired error changes use bootstrap intervals \citep{McNemar1947,Efron1979Bootstrap}, with explicit denominators because oracle Best-of-8 and validation-selected constants differ from deployable single-sample metrics.

\section{Experimental design}
\label{sec:methods}

The audit follows three tiers. Tier 1 (scalar fidelity vs.\ intended pathway) is addressed by the five-group regression comparison and the SMILES-shortcut diagnosis. Tier 2 (direction vs.\ amplitude) is addressed by the target-conditioned direction policy and the two-readout amplitude protocol. Tier 3 (latent control vs.\ decoded structure) is addressed by the P3 input-slot editing and direct final-state injection audits. Unless otherwise stated, 95\% CIs are paired bootstrap intervals and $p$-values are reported as decimals or as $p<0.001$ when smaller.

\subsection{Data, shortcut removal, and evaluation scope}
\label{sec:data_shortcut}

The scalar regression experiments use the aqueous-phase iBonD experimental $\pka$ data as a physical anchor \citep{YangIbonD,Wang2026iBonD}. The principal split contains 9{,}202 training and 1{,}151 held-out test rows. For the $\pka$ task, we remove SMILES from the query while retaining molecular structure through the molecular input channel.

Earlier diagnostics explain why this repair is necessary. In an earlier C configuration with SMILES in the query, \texttt{PkaHead} attained MAE $1.103$ on 24 held-out diagnostic rows, not on the full test set. In the archived modality-conflict summary ($n=24$), replacing only the query SMILES with a donor molecule increased source-molecule MAE from $1.17$ to $4.08$, while donor-molecule MAE became $1.20$; replacing only the SMI-TED features left source MAE at $1.17$ and changed donor MAE only from $4.06$ to $4.07$ (\cref{tab:shortcut,fig:shortcut}). This probe diagnoses a text-driven answer pathway in the pre-repair setting rather than use of the intended molecular channel. After query-SMILES removal, a concurrent token-id alignment repair restored recognition of all six latent slots rather than a single slot; C attains $2.087$ on the 1{,}151-row test set. Because the evaluation cohorts differ and the repairs were not separately ablated, we do not attribute this numerical difference to shortcut removal alone. We therefore use $1.103$ only as a shortcut diagnosis and the repaired setting for all five groups. This is the Tier-1 readout--explanation check: low scalar error need not show that the intended representation drove the computation, anticipating the directional analysis in \cref{sec:direction_amplitude}.

The original split is not strictly scaffold independent: an audit identifies 247 non-empty Murcko scaffolds shared between train and test, affecting 5{,}833 training and 863 test rows; exact canonical-SMILES overlap is zero. Scaffold definitions follow the standard Murcko framework \citep{Bemis1996Properties}. Consequently, the principal regression comparison is an internal architectural evaluation on the original split, not a claim of strict scaffold extrapolation. The frozen-readout experiment applies an additional scaffold-cleaning protocol described below.

\subsection{Model and training}
\label{sec:model_training}

The architecture uses Qwen3-8B-Base as the Transformer backbone \citep{Vaswani2017Attention,Yang2025Qwen3}, with a \emph{prompt} defined as the complete ordered input sequence. Frozen SMI-TED Light features (768 dimensions) are projected by a Q-Former-style cross-attention module with 128 learned queries and eight heads into ChemTokens \citep{Soares2025SMITED,Li2023BLIP2}. Six \texttt{<latent>} positions provide the continuous-inference slots, and the pooled \texttt{<end\_latent>} hidden state is read by \texttt{PkaHead} (layer widths 4{,}096, 2{,}048, 1). Here a \emph{readout} is any function mapping a latent state to a scalar $\pka$ prediction; the main readouts are the nonlinear \texttt{PkaHead} and a separately trained frozen linear ridge readout. We distinguish \emph{latent rollout} (multi-step movement within latent space), \emph{latent intervention} (a directed perturbation of latent states), and \emph{structure editing} (decoded changes to molecular structure). LoRA rank is 16 with $\alpha=32$ and dropout $0.1$; C, B$'$, B$''$, and gen2 train Stage 3 for three epochs. Generation uses temperature $0.9$, top-$p$ $0.95$, seed 0, and at most 96 tokens; P3 additionally reports greedy decoding. Training configurations and budget ablations are summarized in \cref{tab:training_config}; the warm-start exclusion is documented in \cref{sec:training_budget_appendix}. All experiments use a frozen 8B backbone with 43.6M trainable LoRA parameters; the readouts, the policy head, and the generation runs therefore test what a fixed-scale latent interface can and cannot support, not what a larger model might.

\input{tables/table_training_config.tex}

\subsection{Five architectural controls}

\Cref{tab:regression} defines the five groups. A and B evaluate text generation with and without chain-of-thought supervision. B$'$ attaches \texttt{PkaHead} directly to molecular features; B$''$ adds six random positions to control for slot or position effects; C adds the Stage-3 continuous latent-inference pathway before \texttt{PkaHead}. For group $g$, with ground truth $y_i$ and prediction $\hat y_i$, let $e_{g,i}=\lvert\hat y_{g,i}-y_i\rvert$. We report
\begin{equation}
  \MAE_g = \frac{1}{1151}\sum_{i=1}^{1151}e_{g,i},
  \qquad
  \Delta_{A,B}=\frac1n\sum_{i=1}^{n}(e_{A,i}-e_{B,i})=\MAE_A-\MAE_B .
\end{equation}
The same 1{,}151 rows are available for all groups. For the focused C-versus-B$'$ comparison, we resample paired row indices with replacement 10{,}000 times (seed 0), recompute $\MAE_C-\MAE_{B'}$ for each resample, and report the percentile 95\% interval. The two-sided bootstrap sign-tail $p$-value is $2\min(n_{\geq0}+1,n_{\leq0}+1)/(10{,}001)$, capped at one, where the two counts refer to resampled differences above or below zero.

The difference between C and B$'$ can be decomposed as
\begin{equation}
  \MAE_{B'}-\MAE_C
  = \underbrace{\MAE_{B'}-\MAE_{B''}}_{\text{random-position increment}}
  + \underbrace{\MAE_{B''}-\MAE_C}_{\text{remaining latent-path increment}} .
\end{equation}
\begin{equation}
  R=\frac{\MAE_{B''}-\MAE_C}{\MAE_{B'}-\MAE_C}
   =\frac{0.256}{0.276}=0.928 .
\end{equation}
Both the decomposition and $R$ are accounting identities, not a causal variance or information decomposition.

\begin{table}[t]
\centering
\caption{Architectural controls and held-out $\pka$ error after removing the textual SMILES shortcut. All groups are evaluated on the same 1{,}151 test rows. B is nearly constant rather than exactly constant: 1{,}038 predictions equal 1.9 and 113 equal 1.8.}
\label{tab:regression}
\small
\begin{tabular}{p{0.10\linewidth}p{0.55\linewidth}r}
\toprule
Group & Architecture & Test MAE \\
\midrule
A & Text-CoT & 2.895 \\
B & Answer-only text generation (near-constant) & 4.028 \\
B$'$ & Molecular features + \texttt{PkaHead} & 2.363 \\
B$''$ & Molecular features + six random positions + \texttt{PkaHead} & 2.343 \\
\textbf{C} & \textbf{Stage-3 latent inference + \texttt{PkaHead}} & \textbf{2.087} \\
\bottomrule
\end{tabular}
\end{table}

\subsection{Five latent rollout tests}

\paragraph{G1: interpolation.} For pair $i$, we interpolate
$h_i(\alpha)=(1-\alpha)h_{a_i}+\alpha h_{b_i}$ at $\alpha_j=j/20$, $j=0,\ldots,20$, and record $\hat y_i(\alpha_j)=f(h_i(\alpha_j))$. The monotone fraction is
\begin{equation}
  \frac1{38}\sum_{i=1}^{38}\mathbf 1\{\hat y_i(\alpha_0)\leq\cdots\leq\hat y_i(\alpha_{20})
  \ \text{or}\ \hat y_i(\alpha_0)\geq\cdots\geq\hat y_i(\alpha_{20})\}.
\end{equation}
We also report the median jump/span ratio, where
\begin{equation}
 J_i=\frac{\max_{0\leq j<20}\lvert\hat y_i(\alpha_{j+1})-\hat y_i(\alpha_j)\rvert}
 {\max_j\hat y_i(\alpha_j)-\min_j\hat y_i(\alpha_j)} .
\end{equation}
Smooth interpolation is evidence of geometric continuity under this readout, not proof of a true physical manifold.

\paragraph{G2: rollout proximity.} Thirty latent trajectories are optimized for 20 steps toward target $\pka$ values (learning rate $0.005$, target increment $2.0$). For latent state $h$ and the 2{,}302-state background set $\mathcal B$, the pipeline off-manifold score is
\begin{equation}
 d_{\mathrm{off}}(h)=\frac{\min_{h'\in\mathcal B}\lVert h-h'\rVert_2}
 {\operatorname{median}_{h''\in\mathcal B}\lVert h''\rVert_2}.
\end{equation}
The archived rollout stores one score $d_{\mathrm{off}}^{(j)}$ per trajectory, giving $\bar d_{\mathrm{off}}=(1/30)\sum_{j=1}^{30}d_{\mathrm{off}}^{(j)}=0.072$. Target loss is squared error, $L_{j,k}=(f(h_{j,k})-t_j)^2$; we convert it to absolute error as $|f(h_{j,k})-t_j|=\sqrt{L_{j,k}}$ and distinguish medians from the mean final absolute error. Two-dimensional PCA projections visualize the trajectories against $\mathcal B$. The plotting artifact stores trajectory-level scores rather than every raw intermediate state, so G2 remains a supportive pipeline diagnostic, not chemical validation of intermediate states.

\paragraph{G3: latent intervention.} For 40 samples, a target-directed latent intervention is applied through all six latent slots. Let $x_i^{\mathrm{base}}$ and $x_i^{\mathrm{int}}$ denote the complete generated prediction texts, and let $\hat y_i^{\mathrm{base}}$ and $\hat y_i^{\mathrm{int}}$ be the numeric $\pka$ values parsed from those texts. With target $t_i$,
\begin{equation}
 r_i=\mathbf 1\{x_i^{\mathrm{int}}\neq x_i^{\mathrm{base}}\},\qquad
 m_i=\mathbf 1\{\lvert\hat y_i^{\mathrm{int}}-t_i\rvert<\lvert\hat y_i^{\mathrm{base}}-t_i\rvert\}.
\end{equation}
Response therefore means a change in the decoded token sequence, whereas moved requires strictly reduced target error in the parsed value. All 80 archived baseline and intervened predictions lie on a $0.1$ grid, and every text change coincides with a parsed-value change. Consequently, a zero change denotes decoding-level non-response rather than an arbitrarily small continuous readout change. We report $\sum_i r_i/40$ and $\sum_i m_i/40$ separately, together with the distribution of parsed $|\hat y_i^{\mathrm{int}}-\hat y_i^{\mathrm{base}}|$. Additional B$'$ and B$''$ head-input interventions serve as controls in the supplement.

\paragraph{G4: gradient--displacement alignment.} For a molecular pair with latent displacement $d_i=h_{b_i}-h_{a_i}$, directional agreement is
\begin{equation}
  c_i = \frac{\nabla_h f(h_{a_i})^\top d_i}{\lVert\nabla_h f(h_{a_i})\rVert_2\,\lVert d_i\rVert_2},
\end{equation}
where $f$ is \texttt{PkaHead}. We report medians and pair-bootstrap intervals. The three cosine bands used in plots are descriptive conventions only; their definitions, the random-direction reference, and the pair-bank distinction are given in \cref{sec:g4_bands_appendix}.

\paragraph{G5: direction sign.} Let $q_i=\nabla_h f(h_{a_i})^\top(h_{b_i}-h_{a_i})$ be the unnormalized gradient directional derivative. G5 uses $a_i=\mathbf 1\{q_i(y_{b_i}-y_{a_i})>0\}$ and reports $\mathrm{Acc}_{\mathrm{sign}}=(1/105)\sum_i a_i$, with a one-sided exact binomial test of $p=0.5$. This is a coarse sign test and cannot substitute for G4 cosine alignment. Shared molecules can induce pair dependence, so the binomial $p$-value is nominal rather than a fully dependence-adjusted population inference.

\subsection{Target-conditioned direction and amplitude policy}
\label{sec:policy_methods}

For pair $i$, let $\Delta_i=y_{b_i}-y_{a_i}$ and $d_i=h_{b_i}-h_{a_i}$. A two-layer MLP policy $g_\theta(h_{a_i},\Delta_i,y_{a_i})\in\mathbb R^{4096}$ predicts a latent displacement. It is trained on 5{,}456 training-split pairs with
\begin{equation}
 \mathcal L_{\mathrm{dir}}=1-\frac1n\sum_i
 \frac{g_\theta(h_{a_i},\Delta_i,y_{a_i})^\top d_i}
 {\lVert g_\theta(h_{a_i},\Delta_i,y_{a_i})\rVert_2\lVert d_i\rVert_2}.
\end{equation}
Each seed reserves 15\% of training scaffolds for validation and selects the checkpoint by validation median cosine. We use three seeds, hidden width 1{,}024, dropout $0.1$, batch size 128, AdamW with learning rate $10^{-3}$ and weight decay $10^{-4}$, at most 150 epochs, and patience 25. The prespecified direction criterion is median test cosine $>0.3$ on the same 43 held-out pairs used by the expanded G4 bank. The same trained ensemble is additionally re-scored on a pre-registered $151$-pair superset of that bank, defined in the Appendix; no parameter is refitted for the superset. The direction ensemble averages unit-normalized predictions across seeds before measuring cosine.

Amplitude is evaluated separately through the selected frozen ridge vector $\hat w$. With $\hat g_i=g_i/\lVert g_i\rVert_2$, a fixed latent step of size $s$ gives the local readout change
\begin{equation}
 \Delta\hat p_i(s)=s\,\hat g_i^\top\hat w,
 \qquad
 \rho(s)=\operatorname{corr}_i\{\Delta\hat p_i(s),\Delta_i\}.
\end{equation}
We test $s\in\{0.1,0.2,0.3,0.5,1.0,2.0,3.0\}$, reporting median $|\Delta\hat p|$, the fraction in $[0.2,1.0]$, sign accuracy, and signed Pearson correlation. The prespecified amplitude criterion is $\rho>0.5$. A small-step explanation would be supported if some step gave median $|\Delta\hat p|\in[0.2,1.0]$ and $\rho>0.3$; a robust null was predefined as $\rho<0.1$ at every step. This band is a plausibility window for fixed steps, not a calibration target for the full learned norm, whose true median is $3.03$. For scale context, the median true latent displacement norm on these test pairs is $8.21$.

Because $\hat w$ is only one scalar readout, we repeat the same measurements with the nonlinear \texttt{PkaHead} gradient evaluated at the source latent,
\begin{equation}
 \Delta\hat p_i^{\mathrm{head}}(s)=s\,\hat g_i^\top\nabla_h\texttt{PkaHead}(h_{a_i}),
\end{equation}
and with the learned norm, $g_i^\top\nabla_h\texttt{PkaHead}(h_{a_i})$. We also apply the same readout to the true displacement direction $d_i$ as a control. The prespecified decision rule for this reread is recorded in \cref{sec:amplitude_reread_appendix}.

A second run adds explicit amplitude supervision,
\begin{equation}
 \mathcal L=\mathcal L_{\mathrm{dir}}+
 \lambda_{\mathrm{amp}}\frac1n\sum_i\lvert g_i^\top\hat w-\Delta_i\rvert,
 \qquad \lambda_{\mathrm{amp}}=1,
\end{equation}
where $g_i^\top\hat w=\lVert g_i\rVert_2(\hat g_i^\top\hat w)$ is the learned-norm readout change. The joint success criterion is median cosine $\geq0.3$ and learned-amplitude correlation $>0.5$; checkpoint selection remains validation median cosine. The learned-amplitude ensemble averages raw rather than unit-normalized predictions. All ridge-based amplitude measurements in the fixed-step and learned-amplitude policy runs are local frozen-readout proxies, and the explicit amplitude objective optimizes the same ridge projection, so that ablation tests the ridge-targeted objective rather than amplitude control in general. The following P3 analysis separately tests whether the learned direction survives structural decoding.

\subsection{P3: input-slot editing as a control surface}
\label{sec:p3_methods}

\paragraph{Cohort and overlap.} P3 connects the existing policy head to the existing molecular generator rather than introducing a new decoder. Evaluation pairs are constructed only from the 1{,}151 archived held-out states. Following the archived pair-bank rule, candidate pairs require $|\Delta\pka|\geq1$ and Morgan Tanimoto similarity at least $0.6$; each source retains the target whose $|\Delta\pka|$ is closest to $2.0$. Canonical source sorting and deterministic subsampling give 80 eligible unique sources, all of which are evaluated. The 255 generation-training validation rows are not substituted for this cohort. The P3 cohort is not independent of the frozen direction test: all 37 unique sources from its 43 pairs occur among the 80 P3 sources, and 37/43 directed pairs are exactly shared. P3 therefore audits input-slot editing on a largely overlapping policy-test cohort rather than providing an independent held-out replication.

\paragraph{Slot optimization and decoding.} The generator uses its training-matched six-slot coconut prompt. Let $h_i^{\mathrm{base}}$ be its own pooled state at \texttt{<end\_latent>} and $\hat g_i$ the unit-normalized three-seed policy ensemble. The target is $h_i^{\mathrm{target}}=h_i^{\mathrm{base}}+s\hat g_i$ with $s=1.0$. Because the policy operates on pooled $h_T$ rather than directly on the six earlier latent-slot embeddings, only those six embeddings are optimized (Adam, 40 steps, learning rate $0.05$) to minimize relative $h_T$ squared error plus $0.1$ times cosine loss. All backbone and generator parameters remain frozen. Baseline and intervened prompts are decoded with greedy decoding and, separately, multinomial sampling at temperature $0.9$, top-$p$ $0.95$, and seed 0, with at most 96 new tokens. Each sampled baseline/intervention pair resets the same seed to use common random numbers. RDKit canonicalization supplies validity and source-anchored Tanimoto similarity. These P3 metrics are not directly comparable to the DPO direction accuracy, which uses a different cohort and source-relative definition.

Paired learned-oracle scoring, checkpoint-scope caveats, and the initial-slot identity and pass-count diagnostics are summarized in \cref{sec:p3_diagnostics_appendix}.

\subsection{Direct final-hidden-state injection}
\label{sec:hidden_injection_methods}

To separate input-slot optimization from the response at its target tensor, we reuse the same 80 sources, cross-checkpoint policy direction, and decoding protocols as P3. During generation prefill, a forward hook on the backbone's final \texttt{last\_hidden\_state} replaces the final \texttt{<end\_latent>} state by
\begin{equation}
 h_T\leftarrow h_T+s\hat g_i,
 \qquad s\in\{0.1,0.5,1.0,2.0\}.
\end{equation}
No slot or hidden-state optimization is performed. The hook fires once per decode, at the first backbone call; it changes first-token logits through $h_T$ but does not directly edit cached key/value tensors. Each step is paired with its own no-injection baseline under greedy decoding and fixed-seed sampling. We report $\lVert\Delta h\rVert_2/\lVert h_T\rVert_2$, exact sequence and SMILES changes, baseline-anchored Tanimoto similarity, and the learned-oracle direction. The preflight hook check is reported in \cref{sec:hidden_preflight_appendix}.

\subsection{Cross-fitted shared-direction projection}

This exploratory projection over the same 43 test pairs is reported in \cref{sec:shared_projection_appendix}; it is supporting evidence, not independent train--test validation.

\subsection{Frozen ridge readout and geometry}
\label{sec:ridge_methods}

A frozen ridge readout is trained only for scalar accuracy on 2{,}855 scaffold-cleaned fit rows, with $\alpha=100$ selected by validation MAE; 381 fit pairs and 43 test pairs define the directional geometry. The readout solves \citep{Hoerl1970Ridge}
\begin{equation}
 (\hat w,\hat b)=\arg\min_{w,b}\sum_{i\in\mathrm{fit}}(h_i^\top w+b-y_i)^2+\alpha\lVert w\rVert_2^2 .
\end{equation}
We report both original-$h$ and $z$-scored cosines; because cosine is not invariant under dimension-wise standardization, these are sensitivity diagnostics, not interchangeable estimates. Full scaffold-cleaning and geometry details are in \cref{sec:ridge_protocol_appendix}.

\subsection{Readout-free PCA direction test}
\label{sec:pca_methods}

Readout-free PCA fits on the same 2{,}855 rows \citep{Jolliffe2016PCA}. We use the top-90\% projection $P_{90}$ and report test-displacement coverage $r_{90}(d)=\lVert P_{90}d\rVert_2/\lVert d\rVert_2$; a value below $0.2$ is the prespecified low-coverage reference. A separate mean-fit-direction cosine is computed in full space and after projection, so subspace coverage and directional agreement are not conflated. Details are in \cref{sec:pca_protocol_appendix}.

\subsection{Oracle-best SFT, offline DPO, and generation evaluation}
\label{sec:generation_methods}

Generation starts from gen2-final and uses oracle-best candidate selection, $c_i^*=\arg\min_{c\in\mathcal C_i}\lvert\hat p(c)-t_i\rvert$; direction-correct candidates are not required, so this is oracle-best distillation rather than direction-correct-only SFT. Single-sample metrics are direction accuracy and copy rate; Best-of-8 is reported only as an oracle upper bound. Paired gen2-final/SFT comparisons use exact McNemar tests on scored rows, while all-row comparisons use the train-selected constant $3.74$ with missing predictions treated as source/no-op. Offline DPO is initialized from rej1 with $\beta=0.1$; its full construction, optimization, and evaluation protocol are in \cref{sec:dpo_audit}.

\section{Results}
\label{sec:results}

We report results in the order of the three tiers, each a question a world model must answer: does the rollout predict through the molecule (Tier 1), does it know where and how far to move (Tier 2), and does the move survive decoding (Tier 3). Generation is a downstream application of decoded control rather than a fourth independent tier.

\subsection{Tier 1: latent inference, not text, carries the predictive signal}

\begin{figure}[t]
\centering
\includegraphics[width=\textwidth]{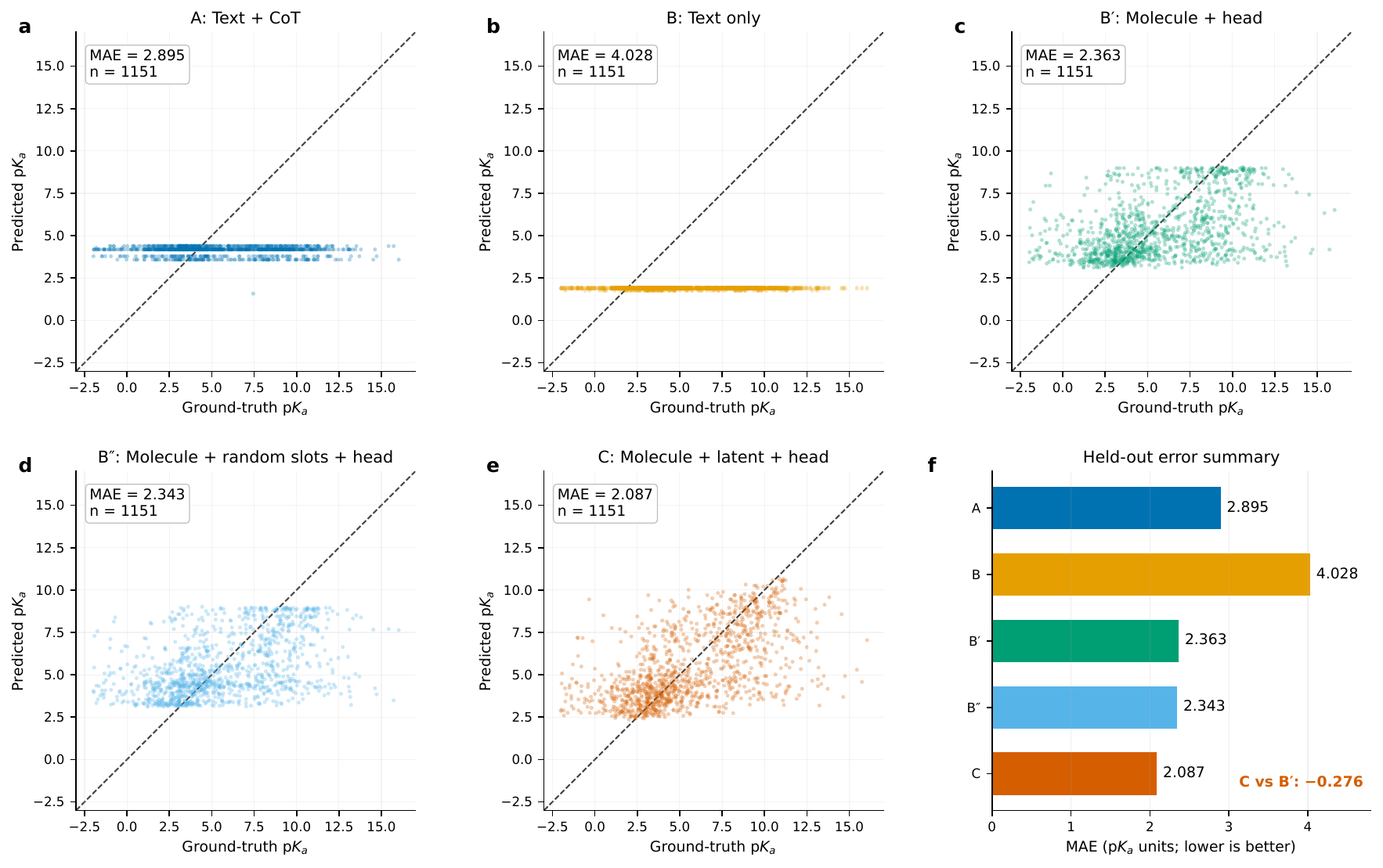}
\caption{Held-out $\pka$ predictions for the five architectural controls on the same 1{,}151 rows. C has the lowest MAE. The focused C-versus-B$'$ comparison has paired MAE difference $-0.2757$ (95\% bootstrap CI $[-0.3525,-0.2029]$, $p<0.001$).}
\label{fig:regression}
\end{figure}

\textbf{Tier 1 asks whether the repaired latent pathway carries predictive signal independent of textual shortcuts.} \Cref{fig:regression} and \Cref{tab:regression} summarize the same 1{,}151 held-out rows. C attains MAE $2.087$, compared with $2.363$ for B$'$, $2.343$ for B$''$, $2.895$ for A, and $4.028$ for B. The paired C-minus-B$'$ MAE difference is $-0.2757$, with 95\% bootstrap CI $[-0.3525,-0.2029]$ and $p<0.001$; all 10{,}000 resampled differences are negative. This supports a paired predictive improvement on the present test split, not external or scaffold-independent generalization. The earlier pre-repair scatter is retained as a developmental baseline in \cref{fig:legacy_indomain}, while \cref{fig:bp_fix_scatter} gives the full repaired B$'$ scatter; neither changes the current five-group comparison.

The B$'\rightarrow$B$''$ change is $0.020$, whereas B$''\rightarrow$C retains a $0.256$ decrease. Thus $0.256/0.276=92.8\%$ of the B$'\rightarrow$C gap remains after random-position control. This percentage is an accounting identity, not a causal variance or information decomposition.

Answer-only generation is nearly constant and has the largest MAE, while Text-CoT improves over it but remains above the molecular-head models. The comparison therefore isolates predictive differences within the repaired pipeline rather than reproducing the earlier SMILES-reading shortcut. Subsequent tests measure geometry, intervention, and control directly rather than inferring them from MAE.

\subsection{Latent geometry is smooth, responsive, and stream-consistent}

\begin{figure}[t]
\centering
\includegraphics[width=\textwidth]{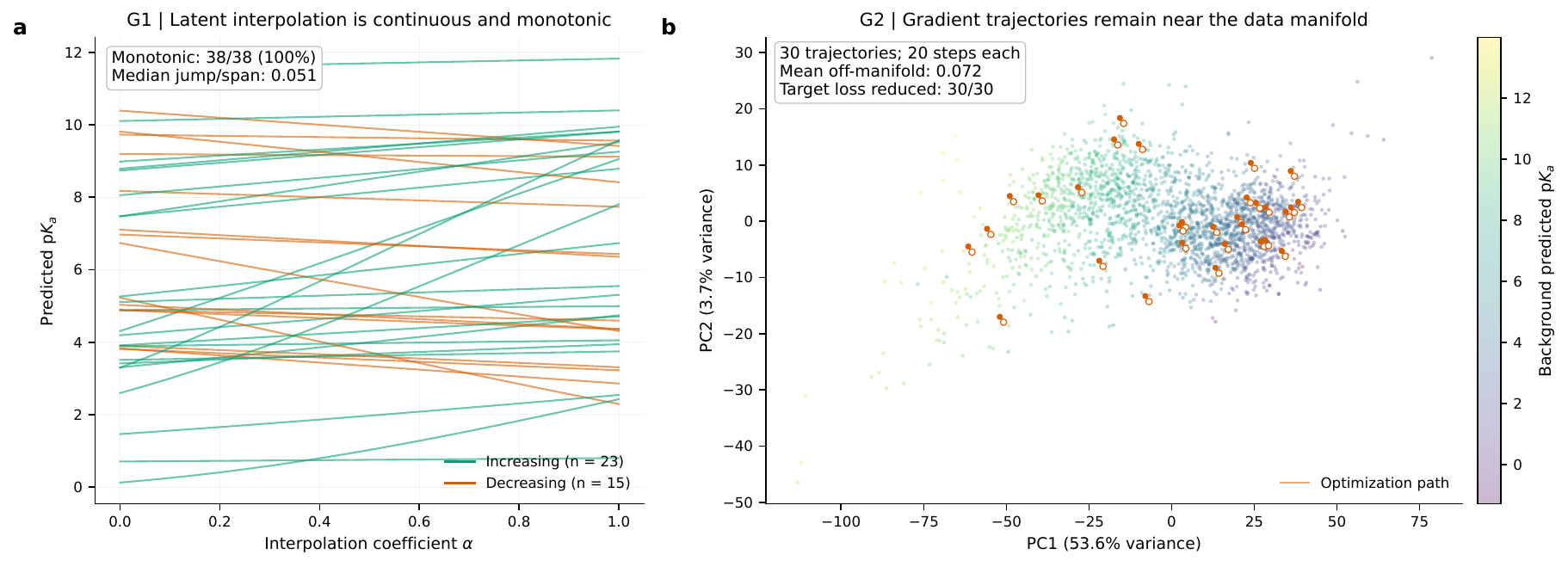}
\caption{Latent interpolation and rollout. (a) All 38 pairwise interpolations are monotonic under \texttt{PkaHead}. (b) Thirty 20-step target-directed trajectories reduce target loss; their mean off-manifold score is $0.072$. The PCA projection is a visualization, not evidence that two coordinates capture the full latent geometry.}
\label{fig:latent_structure}
\end{figure}

All 38 G1 interpolation curves are monotonic, with a median jump/span ratio of $0.051$ (\Cref{fig:latent_structure}a). Across 30 G2 optimization trajectories, all 30 reduce target loss and the mean off-manifold score is $0.072$ (\Cref{fig:latent_structure}b). The archived quantity is squared target loss: its median falls from $4.000$ to $0.0631$. Taking the square root gives median absolute target error $2.000\rightarrow0.251$; the final mean absolute error is $0.288$ (\cref{fig:g2_loss_conversion}). These results support continuous readout behavior and multi-step error accumulation under the measured criteria. They do not prove that every interpolated or optimized state corresponds to a realizable molecule with a physically valid $\pka$. The original trajectory visualizations are described in \cref{sec:legacy_artifacts}; numerical claims use the current G1/G2 metrics rather than the legacy summary annotations embedded in those graphics.

\begin{figure}[t]
\centering
\includegraphics[width=\textwidth]{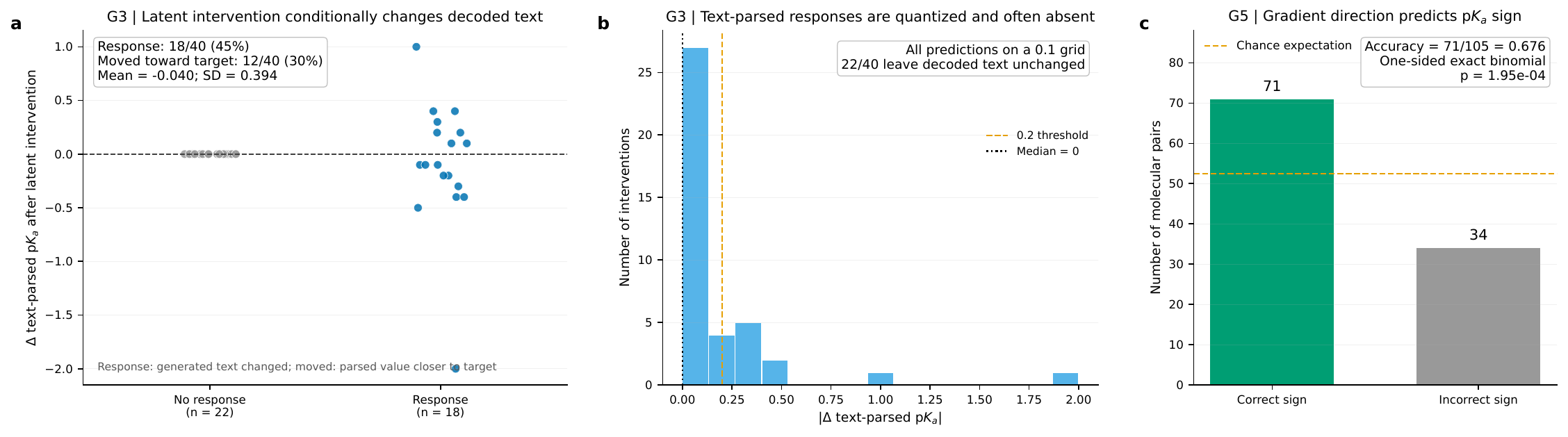}
\caption{Conditional intervention, text-parsed response, and coarse sign prediction. (a) Eighteen of 40 latent interventions change the generated prediction text; 12 move toward target. (b) Twenty-two interventions leave the decoded text and parsed value unchanged. All parsed predictions lie on a $0.1$ grid, so zero denotes decoding-level non-response rather than an arbitrarily small continuous change. (c) G5 sign accuracy is $0.676$ (71/105). G3 and G5 answer different questions; neither implies calibrated continuous amplitude control.}
\label{fig:intervention}
\end{figure}

G3 directly tests intervention response at the decoded-text level. Eighteen of 40 interventions change both the generated text and its parsed prediction, and 12 move closer to target; the remaining 22 leave both unchanged (\Cref{fig:intervention}a). The mean signed parsed change is $-0.040$ with standard deviation $0.394$. Because every parsed prediction lies on a $0.1$ grid, the median zero change means that half of the interventions do not cross the decoding threshold, not that a continuous output moves by an arbitrarily small amount (\Cref{fig:intervention}b). This discrete non-response is distinct from the policy head's continuous amplitude correlation and does not by itself establish continuous amplitude control in either direction. The 22 non-responsive cases are not explained solely by small target gaps or absence of ionizable groups (\Cref{tab:nonresponse}). Additional zero, swap, and head-input controls are presented in \cref{fig:s1,fig:s2,fig:s3,fig:s4}.

G5 reaches $71/105=0.676$ sign accuracy (one-sided exact binomial $p=1.95\times10^{-4}$; \Cref{fig:intervention}c). Together, G1--G2 support continuous readout behavior and multi-step accumulation, G3 shows partial response at the decoded-text level, and G5 shows that the local gradient carries coarse sign information.

\begin{table}[t]
\centering
\caption{Four supporting tests of latent geometry, intervention, and directional sign. Entries labeled Pass or Conditional refer to prespecified thresholds for G1--G3 and G5. G2 distinguishes squared loss from absolute error; G3 reports decoded-text response alongside the text-parsed numeric distribution. Naive G4 gradient alignment is retained as a supporting diagnostic in \cref{fig:g4} rather than as the main bottleneck.}
\label{tab:gates}
\small
\begin{tabular}{p{0.29\linewidth}p{0.43\linewidth}p{0.16\linewidth}}
\toprule
Gate & Result & Reading \\
\midrule
G1 interpolation monotonicity & 1.000 (38/38) & Pass \\
G2 rollout accumulation & median $\lvert\mathrm{error}\rvert$ $2.000\rightarrow0.251$; off-manifold $=0.072$ & Pass \\
G3 latent intervention & text response $=18/40$; moved $=12/40$; unchanged $=22/40$ & Conditional \\
G5 direction sign & 0.676 (71/105) & Pass \\
\bottomrule
\end{tabular}
\end{table}

\begin{table}[t]
\centering
\caption{Diagnosis of the 22 non-responsive C-group interventions. The smaller target gap and higher ionizable-group fraction in the non-responsive subset are inconsistent with chemical inaccessibility as the sole explanation. Ionizable fraction is the archived pipeline annotation; its underlying functional-group rule is not preserved in the plotting artifact and is therefore not reconstructed here.}
\label{tab:nonresponse}
\small
\begin{tabular}{lcccc}
\toprule
Subset & n & Median $\lvert$target$-$base$\rvert$ & Fraction $>2.0$ & Ionizable fraction \\
\midrule
Responsive & 18 & 2.79 & 0.72 & 0.61 \\
Non-responsive & 22 & 2.42 & 0.59 & 0.77 \\
\bottomrule
\end{tabular}
\end{table}

\subsection{Tier 2: direction is learnable, amplitude is readout-dependent}
\label{sec:direction_amplitude}

Having established that the latent pathway carries predictive signal (Tier 1), we now test the core hypothesis of Tier 2: direction and amplitude are separable capabilities, and amplitude estimates depend fundamentally on the choice of scalar readout. We first verify that the target-conditioned policy can learn displacement directions, then show that the same policy yields diametrically opposite amplitude conclusions under two different readouts, and finally test whether explicit amplitude supervision can resolve the discrepancy.

\subsubsection{Target-conditioned policies reliably learn latent displacement directions} The three target-conditioned policy seeds attain median test cosines $0.482$, $0.539$, and $0.469$; the median of seed medians is $0.482$, and every seed passes the prespecified $>0.3$ criterion. The unit-normalized ensemble reaches $0.577$ (\Cref{fig:direction_amplitude}a and \cref{tab:direction_amplitude}). We treat the pair fraction as the primary policy statistic because it is non-parametric and does not inherit the low median interval bound: $27/43=0.628$ pairs exceed cosine $0.3$ for seeds 0 and 2 and for the ensemble (exact 95\% CI $[0.467,0.770]$; one-sided exact binomial $p<0.001$ against a $0.3$ rate), and $26/43=0.605$ for seed 1. The ensemble median interval lower bound is $0.235$, so the point estimate alone would not clear the criterion robustly; the pair count does. Validation curves peak at epochs 28, 43, and 21, followed by early stopping at epochs 53, 68, and 46 under the fixed patience of 25; the rerun reproduces all three best epochs and test medians exactly (\cref{fig:policy_validation_curves}). This rules out stopping at the 150-epoch cap as the explanation, without claiming that every possible longer-training variant would be equivalent. These results confirm that directional information is explicitly learnable from paired molecular data, and that the policy captures meaningful geometric structure beyond naive gradient alignment. This does not, however, imply that the same direction carries a predictable property amplitude---a question we turn to next.

\subsubsection{Held-out bank expansion reproduces direction and does not rescue amplitude} The 43-pair bank is small enough that its interval bound is weak, so we re-scored the same trained policy on a pre-registered $151$-pair superset, whose rule is fixed in the Appendix. The rule relaxes the Tanimoto floor from $0.6$ to $0.5$, the value already used for the training pair bank, and otherwise leaves the pair definition unchanged, so the original $43$ are retained unchanged and $108$ pairs with $0.5\leq\mathrm{Tanimoto}<0.6$ are added. Training pairs, seeds, epoch budget, and checkpoint selection are untouched; only the evaluation file changes, and re-running the original bank reproduces the ensemble median $0.577$ exactly.

Direction is unchanged by the expansion. The ensemble median is $0.547$ on the pooled $151$ pairs (pair-bootstrap 95\% CI $[0.407,0.625]$), against $0.577$ ($[0.235,0.695]$) on the original $43$, and the new $108$ pairs alone give $0.543$ ($[0.403,0.626]$). The pair fraction is nearly identical: $95/151=0.629$ exceed cosine $0.3$, against $27/43=0.628$ on the original bank and $68/108=0.630$ on the new pairs. The interval width therefore roughly halves while the point estimate moves by $0.03$, so the direction result is not an artifact of the original similarity floor. The $108$ added pairs are disjoint from the original $43$ as pairs, and $79$ of them also share no molecule with the original bank. That strictest subcohort gives median $0.435$ (95\% CI $[0.257,0.605]$) with $59.5\%$ above cosine $0.3$: the median falls, as expected once every shared molecule is removed, while the pair fraction still clears the criterion and its interval lower bound does not.

The amplitude verdict is stable but its precise boundary is bank-dependent. Under the frozen ridge readout the signed correlation at a fixed step is $-0.020$ on the original $43$ pairs and $0.132$ on the expanded $151$; the learned-norm correlation is $0.014$ and $0.138$; sign accuracy is $0.395$ and $0.470$. Every step size shares the expanded value because Pearson correlation is invariant to positive rescaling. All of these remain below both the prespecified calibration condition $\rho>0.3$ and the success criterion $\rho>0.5$, so the amplitude conclusion holds. The stronger statement that $\rho<0.1$ at every step, however, is specific to the 43-pair bank and does not survive expansion, and we do not carry it over. The expanded bank raises power over the same $1{,}151$ held-out states; it is not an independent external cohort.

\subsubsection{The ridge amplitude null is a geometric artifact of the readout, not evidence of no amplitude signal} The selected frozen ridge has a median cosine of $1.18\times10^{-5}$ to the true displacement directions, effectively orthogonal to them (Section~\ref{sec:ridge_methods}). For such a readout, the quantity $\hat g_i^\top\hat w$ has almost no target-relevant variance by construction, so a near-zero amplitude correlation is expected whatever the policy does. A near-zero amplitude correlation is therefore a predetermined consequence of the readout's geometry, not a neutral measurement of the policy's amplitude control. We therefore repeated every amplitude measurement with the nonlinear \texttt{PkaHead} gradient, whose median direction cosine to the same displacements is $0.107$ rather than zero (Section~\ref{sec:policy_methods}).

\subsubsection{Readout geometry, not fitting quality, decides whether amplitude is recoverable} We next asked whether the ridge null is a property of that one readout or of independently fitted readouts in general. On the same frozen latents ($9{,}202$ train, $1{,}151$ validation, $1{,}151$ test states; \cref{sec:readout_family_methods}), we fitted nine scalar readouts: five ridge penalties ($\alpha=0.01$--$100$), two one-hidden-layer ReLU heads ($64$ and $256$ units), a label-shuffled ridge, and a half-data ridge. Each readout was scored by its input gradient $g=\nabla_h f$ at the source state and by the amplitude it assigns to a displacement, $\Delta=f(h_b)-f(h_a)$.

\textbf{Proposition.} Let $w$ be a scalar readout's input gradient and $d$ a latent displacement. The predicted property change is $\langle w,d\rangle=\lVert w\rVert\lVert d\rVert\cos(w,d)$. Amplitude ordering across pairs is therefore recoverable only if $\lvert\cos(w,d)\rvert$ is non-negligible on the displacement set. For a readout fitted independently of the pathway, $w$ carries no preferred relation to $d$; for fixed $d\in\mathbb{R}^{D}$, $\cos(w,d)$ concentrates at scale $D^{-1/2}$, which at $D=4096$ is $0.0156$. Geometric blindness is thus the default state of an independently fitted readout, not a symptom of underfitting.

\begin{figure}[t]
\centering
\includegraphics[width=\textwidth]{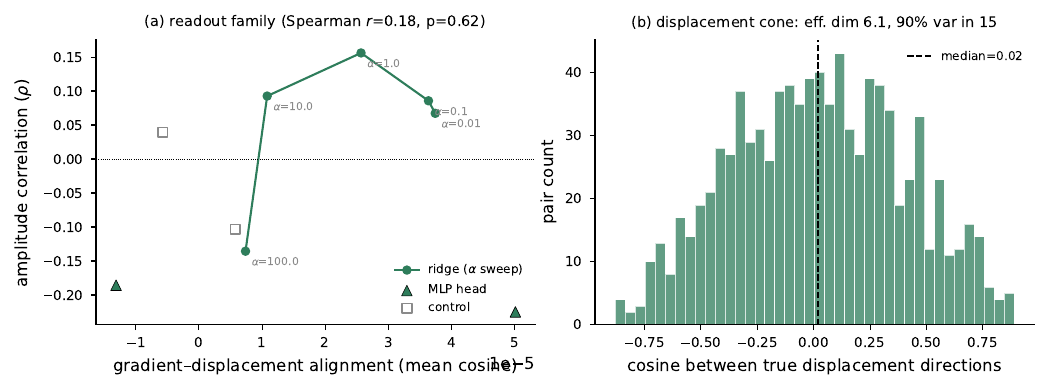}
\caption{Readout geometry and the displacement cone. (a) Nine readouts fitted on the same frozen latents: gradient--displacement alignment (abscissa) against amplitude correlation (ordinate). Independently fitted readouts cluster at alignment $\lesssim5\times10^{-5}$, three orders of magnitude below the random-direction scale $1/\sqrt{4096}=0.0156$, and none recovers amplitude; the label-shuffled and half-data controls overlap the fitted heads. (b) Distribution of pairwise cosines among the 43 true displacement directions; the set concentrates in few effective dimensions (participation ratio $6.1$; $90\%$ of variance in $15$ dimensions) without a shared mean axis.}
\label{fig:readout_family}
\end{figure}

All nine readouts behave as the proposition predicts. Mean gradient--displacement alignment lies between $5.7\times10^{-6}$ and $5.0\times10^{-5}$, at most $0.32\%$ of the random-direction scale, and no readout produces a non-null amplitude correlation ($\rho$ from $-0.225$ to $+0.156$; all $p\geq0.14$; \cref{fig:readout_family}a). Label-shuffled ($\rho=0.039$) and half-data ($\rho=-0.103$) controls are indistinguishable from the fitted heads, and the $256$-unit head is no better than the $64$-unit head ($\rho=-0.186$ versus $-0.225$), so additional capacity does not help. The only readout in this study with a non-negligible alignment is the pathway-co-trained \texttt{PkaHead} gradient ($0.107$; Section~\ref{sec:direction_amplitude}), roughly three orders of magnitude above the independently fitted family, and it is also the only one that recovers relative amplitude order ($\rho=0.585$).

The displacements themselves occupy a narrow, low-dimensional set (\cref{fig:readout_family}b): $90\%$ of their variance lies in $15$ of $4{,}096$ dimensions, with a participation ratio of $6.1$, while the directions are not mutually aligned (median pairwise cosine $0.020$). The best ridge gradient makes an angle of $90.0^\circ$ with the mean displacement axis. Readout choice therefore does not merely change the amplitude number; it decides whether the amplitude question is well posed. Scalar fit quality is uninformative about this geometry: validation MAE moves from $2.47$ to $1.85$ across the ridge sweep while alignment changes by a factor of five at $10^{-5}$ magnitude.

\paragraph{Where the readout weight actually lives.} A natural mechanism is that least squares concentrates $w$ in the high-variance subspace that the displacement cone avoids. The archived fit supports neither half of this account. The cone is not low-variance: $63\%$ (43 pairs) and $67\%$ (151 pairs) of the unit displacements' energy lies inside the six leading principal components that carry $90.2\%$ of fit variance. The selected ridge is not spread across that subspace either---only $4.9\times10^{-9}$ of $\lVert w\rVert^2$ lies in it, against $1.5\times10^{-3}$ for a random direction. Instead the readout degenerates onto a single coordinate carrying $98.5\%$ of $\lVert w\rVert^2$ ($99.99\%$ for the top two), and that coordinate is nearly orthogonal to the cone. Removing those two coordinates leaves a weight statistically indistinguishable from a random direction: it captures $0.50\%$ of its energy in the 43-pair cone against a $1.05\%$ random reference, and $3.8\%$ against $3.4\%$ on the 151-pair bank. The blindness is therefore a degeneracy of an independently fitted scalar readout rather than an active steering away from the cone. We offer this as an interpretation of archived weights, not a verified mechanism: it is measured on the selected ridge, and although the other eight readouts show the same blindness, we did not test whether they share the degeneracy.

\subsubsection{Larger fixed steps do not rescue ridge amplitude} 
\begin{figure}[t]
\centering
\includegraphics[width=\textwidth]{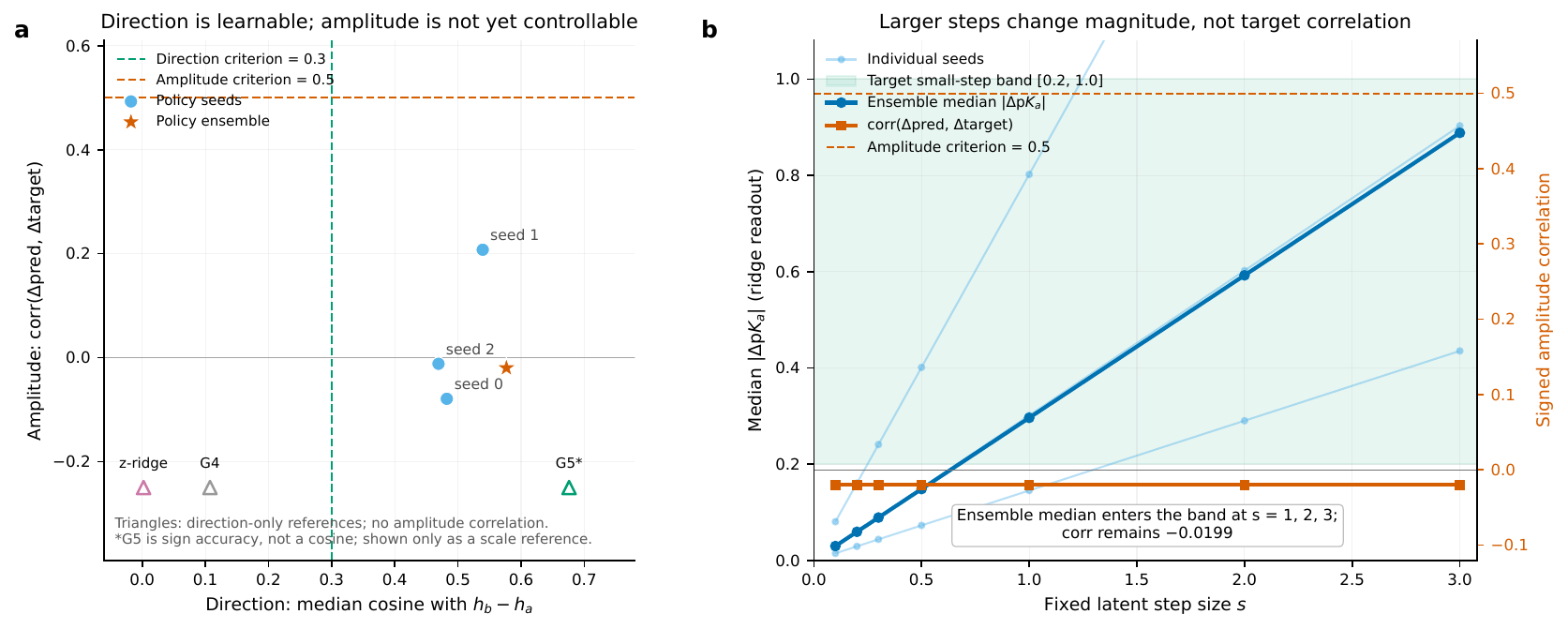}
\caption{Direction and amplitude on 43 held-out pairs. (a) All three policy seeds and their unit-normalized ensemble pass the direction criterion (median cosine $>0.3$). The amplitude correlations plotted here use the frozen ridge readout, whose median cosine to true displacements is $1.18\times10^{-5}$; the nonlinear \texttt{PkaHead}-gradient reread is reported in \cref{tab:direction_amplitude} and in the text. G4 and $z$-scored ridge are direction-only references without comparable target-conditioned amplitude measurements. G5 is sign accuracy, not cosine, and is shown only as a scale reference. (b) Increasing fixed latent steps changes the frozen-ridge readout magnitude. Ensemble medians enter $[0.2,1.0]$ at $s=1,2,3$, while ridge correlation remains $-0.0199$.}
\label{fig:direction_amplitude}
\end{figure}

For the same ensemble, fixed steps $s=0.1,0.2,0.3$ produce median $|\Delta\hat p|$ of $0.0296$, $0.0592$, and $0.0888$, respectively. The much larger steps $s=1,2,3$ produce $0.296$, $0.592$, and $0.888$, placing the ensemble median inside the target small-step band $[0.2,1.0]$ (\Cref{fig:direction_amplitude}b). For scale context, the median true latent displacement norm is $8.21$. Nevertheless, the ridge signed amplitude correlation remains $-0.0199$ at every positive step size: all steps satisfy the prespecified robust-null condition $\rho<0.1$, and none meets the calibration condition $\rho>0.3$ or the amplitude success criterion $\rho>0.5$. Pearson correlation is invariant to positive fixed rescaling, so this test separates merely achieving a nonzero target-range magnitude from predicting the requested magnitude under this readout.

\subsubsection{A nonlinear readout recovers relative amplitude order} With $\nabla_h\texttt{PkaHead}$ evaluated at the source latent, the ensemble gives signed correlation $0.460$ at $s=1$ (pair-bootstrap 95\% CI $[0.237,0.629]$; molecule-cluster CI $[0.180,0.647]$) and $0.585$ for the learned norm (pair CI $[0.254,0.798]$; cluster CI $[0.215,0.811]$). The learned-norm correlation exceeds the prespecified $0.5$ threshold while the ensemble cosine is $0.577$, so the joint criterion is met; all three seeds also meet it under this readout ($0.561$, $0.523$, and $0.538$). Sign accuracy is $0.744$--$0.767$.

The absolute scale is compressed rather than absent. The learned norm has median $2.33$ ($3.11$ in absolute value) against a true median $|\Delta|$ of $3.03$; the median predicted-to-true ratio is $0.77$, only $40\%$ of pairs fall within a factor of $1.5$, and $11/43$ carry the wrong sign. A fixed unit step predicts $0.047$, but that step is about $68$ times shorter than the policy's own norm, so its small value is expected and is not a scale-calibration test. The learned-norm correlation is not attributable to direction alone: $\|g_i\|$ correlates $0.355$ with $\Delta_i$, the unit-direction term $\hat g_i^\top\nabla f$ correlates $0.460$, and their product correlates $0.585$.

A single global factor does not repair the scale. Fitting one multiplicative constant on the 5{,}456 training pairs gives $\sigma=0.217$ by least squares and $0.821$ by the median ratio; applied unchanged to the held-out pairs, these predict medians of $0.51$ and $1.91$ against a true $3.03$, with MAE $2.98$ and $2.35$ and only $9\%$ and $37\%$ of pairs within a factor of $1.5$. Even the oracle in-sample factor ($\sigma=0.601$) leaves MAE $2.37$ and $26\%$ within a factor of $1.5$. The true-displacement control gives $\rho=-0.093$ (95\% CI $[-0.407,0.289]$), so the nonlinear readout does not recover amplitude for the data's own displacement directions either. The two readouts therefore disagree and neither alone identifies physical amplitude control: the ridge null is mechanically forced, while the \texttt{PkaHead} correlation shows that the policy direction and step length both track the requested change without reproducing a calibrated magnitude.

\input{tables/table5_direction_amplitude.tex}

\subsubsection{Ridge-targeted amplitude supervision does not restore joint control} With $\lambda_{\mathrm{amp}}=1$, the learned-norm ensemble has median cosine $0.192$ and ridge amplitude correlation $0.126$, again failing the joint criterion. Seed 2 attains correlation $0.809$ but cosine $0.232$; seed 0 attains cosine $0.312$ but correlation $-0.320$; no individual seed satisfies both criteria (\cref{tab:amplitude_supervision}). The reciprocal seed-level pattern is consistent with an objective trade-off in this tested setting, rather than a simple account in which both metrics are uniformly undertrained; it does not establish a universal mechanism. Because this run optimizes the ridge projection that the reread shows to be nearly orthogonal to displacements, its null result applies to that targeted objective and not to amplitude supervision in general. Multi-objective checkpoint selection or staged training is therefore a next step, not a result established here. G3 provides a separate decoding-side observation: 22/40 interventions leave the generated prediction text unchanged. This is not a continuous-amplitude measurement.

\subsubsection{Supporting directional diagnostics} Naive \texttt{PkaHead} gradients give median G4 cosine $0.107$ with pair-bootstrap CI $[-0.001,0.175]$ (\cref{fig:g4}). The original small-gap summary $0.064$ uses $n=6$, whereas the expanded-bank summary $0.022$ uses $n=13$; these are different pair banks and are not comparable as repeated estimates. Exploratory cross-fitted shared-direction projections give median $0.244$ (95\% CI $[-0.078,0.386]$) and a mean-based sign-flip $p=0.0150$ (\cref{fig:s8}). The interval and exploratory pair reuse keep this as supporting rather than definitive population evidence.

The selected frozen ridge readout reaches test MAE $2.259$, compared with $2.087$ for the existing head and $2.882$ for the train-mean baseline, yet its original-$h$ median cosine is $1.18\times10^{-5}$ and its $z$-scored median is $0.00217$ (95\% CI $[-0.00047,0.00521]$; \cref{fig:s9}). Scalar fit alone therefore does not recover the displacement direction for this selected readout. Two coordinates carry 99.99\% of the original-$h$ weight energy (exact fraction $0.999878$), motivating the separate $z$-scored sensitivity geometry rather than a direct comparison of unlike cosines.

Readout-free PCA further argues against a primarily low-variance explanation here: six components capture $90.19\%$ of fit variance, and median test-displacement coverage is $0.818$ (95\% CI $[0.772,0.849]$; \cref{fig:s10}). Mean-fit-direction transfer has median cosine $0.284$ before and $0.282$ after projection, with both intervals crossing zero. Together, the diagnostics demote naive gradient alignment to a weak baseline and localize the central result at the policy level: direction passes its criterion, whereas the amplitude conclusion depends on which readout is used. Because the same 43 pairs enter G4, the policy, ridge, PCA, shared projection, and the amplitude reread, all reported $p$-values are nominal and no multiplicity correction is applied.

\subsection{Tier 3a: input-slot editing does not isolate a decoder bottleneck}
\label{sec:p3_results}

\textbf{Tier 3 audits the transferability of latent control to decoded molecular structures.} A latent displacement that changes scalar readout values does not necessarily alter the token-level generation pathway enough to produce a different molecular structure. We test two distinct control surfaces---input-slot optimization and direct final-hidden-state injection---to bound where control breaks down in the current architecture. Under greedy decoding, optimizing only the six slot embeddings realizes the requested pooled state almost exactly: median $h_T$ target cosine is $0.99997$ and median relative error is $0.00807$ across all 80 sources. Structural decoding nevertheless changes only $1/80$ output sequences. The intervened SMILES are all valid, but median source-anchored Tanimoto similarity is $0.576$, below the $0.6$ check. The paired intervened-minus-baseline learned-oracle effect has direction accuracy $0/80$ and median $|\Delta\pka|=0$; 79 cases have exactly zero effect, and the sole changed sequence moves opposite to its requested direction. Thus the Tanimoto, direction, and amplitude checks fail despite successful pooled-state optimization, whereas validity passes (\cref{tab:p3_decode}). We read this as an unsuccessful cross-checkpoint input-slot editing attempt rather than as a localized decoder bottleneck. The identity check is limited to the initial pre-optimization input embeddings; their near-invariance neither bounds the optimized displacement nor shows that the states used during decoding are molecule-invariant. P3 therefore tests the input-slot control surface; \cref{sec:hidden_injection_results} separately tests direct final-state injection. The token-repair control (\cref{fig:s4}; $19/24$ under zeroing) shows that a large slot perturbation can alter generated sequences, but it does not establish target-directed control.

The fixed-seed sampled protocol gives the same median target-state cosine and relative error but changes $0/80$ sequences (\cref{tab:p3_decode_sample}). Intervened validity is $75/80=0.9375$, passing its check, while median source-anchored Tanimoto is $0.514$, again below $0.6$. Seventy-five pairs have both learned-oracle scores; the remaining five identical invalid pairs are retained as zero effects by output identity. Paired direction is $0/80$, all 80 effects are zero, and median $|\Delta\pka|$ is zero. Sampling at the generation-matched temperature therefore provides no evidence that the absence of changes under greedy input-slot editing is merely a decoding-mode artifact, although one paired sampling seed does not characterize variability across seeds.

\subsection{Tier 3b: final-state injection does not change decoding at the tested norms}
\label{sec:hidden_injection_results}

The median relative displacements $\lVert\Delta h\rVert_2/\lVert h_T\rVert_2$ are $0.000675$, $0.004160$, $0.008178$, and $0.016170$ for $s=0.1,0.5,1.0,2.0$, respectively; the median baseline norm is $123.896$. At $s=2$, the per-source relative displacement ranges from $0.007109$ to $0.016660$. All eight step-by-protocol arms change $0/80$ token sequences and $0/80$ parsed SMILES relative to their paired baselines. Baseline-to-intervention Tanimoto similarity is $1.0$ for every pair in which both structures are valid (80 greedy pairs and 75 sampled pairs). Intervened validity is $1.0$ under greedy decoding and $0.9375$ under sampling. No learned-oracle direction successes occur, but this follows from exact output identity and is not evidence that the policy direction is incorrect.

The preflight logit change confirms an active injection path, so the null is not explained by a disconnected hook. It shows only that adding $s\hat g$ once at the final \texttt{<end\_latent>} state is insufficient to alter decoding over the realized range. Larger perturbations, other layers or positions, persistent key/value-cache interventions, and ChemToken editing remain separate tests; this experiment does not establish a decoder-wide bottleneck.

\subsection{Tier 3c: generation improves over the prior generator but stays bounded}

Under the preregistered rule, the one-epoch SFT and two-epoch DPO results are primary; the three- and four-epoch results are sensitivity ablations because neither changes the qualitative conclusion.

\begin{figure}[t]
\centering
\includegraphics[width=\textwidth]{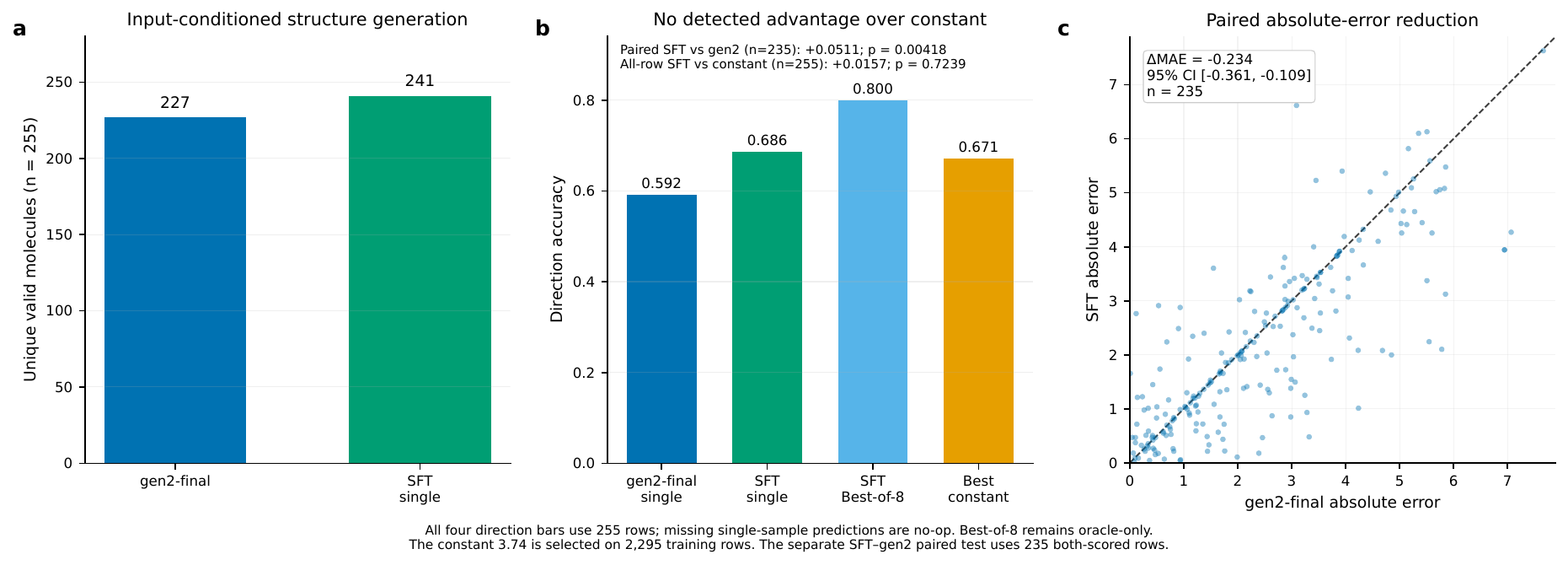}
\caption{Generation-path evaluation. (a) Unique valid structures for gen2-final and SFT single-sample generation. (b) Direction accuracy for gen2-final, SFT single, SFT Best-of-8, and the train-selected constant $3.74$. All four bars use 255 rows with missing single-sample predictions treated as no-op; Best-of-8 is an oracle upper bound. (c) Paired absolute errors on the 235 both-scored rows show a mean SFT-minus-gen2-final change of $-0.234$ (95\% CI $[-0.361,-0.109]$).}
\label{fig:generation}
\end{figure}

The generation path instantiates latent rollout as target-conditioned structure editing. Gen2-final already generates input-conditioned structures; SFT does not create this ability from scratch. At the primary one-epoch budget, it increases validity from $0.922$ to $0.988$, unique valid structures from 227 to 241, and mean Tanimoto similarity from $0.556$ to $0.581$ (\Cref{fig:generation}a). The copy rate simultaneously increases from $0.051$ to $0.122$.

On the 235 both-scored rows, one-epoch SFT improves direction accuracy from $0.6426$ to $0.6936$. Fourteen rows become correct only under SFT and two become correct only under gen2-final, giving exact McNemar $p=0.00418$. Paired MAE decreases by $0.234$, with a 95\% bootstrap CI of $[-0.361,-0.109]$ (\Cref{fig:generation}c). This is the \emph{paired generation improvement}: a target-directed gain over the previous generator under learned-oracle evaluation, not evidence of experimentally validated physical causation. At three epochs, on the 225 both-scored rows, the same comparison gives $0.6400\rightarrow0.6800$ with 6 gen2-only and 15 SFT-only rows (exact McNemar $p=0.0784$), so the direction gain is no longer significant; the paired MAE change remains $-0.252$ (95\% CI $[-0.404,-0.098]$).

On all 255 rows with missing predictions treated as no-op, the train-selected constant $3.74$ attains $171/255=0.6706$ direction accuracy and one-epoch SFT attains $175/255=0.6863$. Thirty-four rows are correct only for the constant and 38 only for SFT, giving exact McNemar $p=0.7239$. At three epochs, SFT attains $166/255=0.6510$ with 41 constant-only and 36 SFT-only rows ($p=0.6488$). An advantage over the constant is therefore not detected at either budget; non-significance must not be read as equivalence. This all-row constant comparison is separate from the paired SFT--gen2-final test above.

Best-of-8 reaches direction accuracy $0.800$ and MAE $1.313$ at one epoch and $0.780$ and $1.414$ at three epochs, providing an explicit oracle upper bound on the target-directed improvement available within the candidate set. It is not a single-sample result or a practical selection mechanism. The longer SFT budget also degrades decoded-output quality: validity falls from $0.988$ to $0.949$, unique valid structures from 241 to 233, mean Tanimoto from $0.581$ to $0.532$, and scored rows from 252 to 242, while the copy rate falls from $0.122$ to $0.078$. The no-op reference attains target MAE $2.293$ over all 255 rows and has no direction prediction under the reporting convention. Together, the copy-rate decrease and comparatively strong no-op MAE are consistent with more conservative editing after longer SFT, but they do not establish conservatism as the causal mechanism of improvement. Across these two evaluated budgets, one epoch has the stronger direction-accuracy and decoded-output-quality profile; three epochs reduces decoded-output quality without improving direction accuracy. This two-point comparison does not identify a general optimal budget. The pattern is consistent with early overfitting, but the archived runs do not include a held-out loss curve that would establish the mechanism.

We next test whether offline preference optimization internalizes target-directed generation. Starting from rej1, two-epoch DPO attains $182/255=0.7137$ single-sample direction accuracy with missing predictions treated as source/no-op, compared with $173/255=0.6784$ for rej1, but remains below the $0.72$ design reference. The train-selected constant $3.74$ attains $171/255=0.6706$; 28 rows are correct only for the constant and 39 only for DPO, giving exact McNemar $p=0.2215$. The supplementary direct DPO--rej1 comparison also does not detect an improvement (14 rej1-only and 23 DPO-only rows, $p=0.1877$). Copy rate falls to $1/255=0.0039$, below the $0.15$ design bound. Doubling the budget to four epochs does not rescue the criteria: direction accuracy falls to $162/255=0.6353$, below both the two-epoch result and the constant, with 45 constant-only and 36 DPO-only rows ($p=0.3742$); the direct rej1 comparison remains non-significant (29 rej1-only and 18 DPO-only rows, $p=0.1439$). The primary McNemar test therefore does not detect an advantage over the constant at either budget, and the numerical accuracy differences do not establish a generator improvement.

At two epochs, nine DPO rows without a scored prediction are treated as source/no-op; at four epochs, 29 rows are, while scored rows fall from 246 to 226. All rows are retained in the 255 denominators. Two-epoch DPO Best-of-8 gives $0.7647$ under this all-row convention ($0.7927$ among 246 scored rows) and four-epoch DPO gives $0.6784$ ($0.7655$ among 226 scored rows), strictly as an oracle diagnostic. No post-result tuning or criterion relaxation is performed. Full pair-construction and optimization details are given in \cref{sec:dpo_audit}.

\input{tables/table4_generation.tex}

\section{Discussion}
\label{sec:discussion}

\subsection{Direction and amplitude are distinct capabilities on the path to causal world models}

Our central finding is that directional alignment and amplitude calibration are not two sides of the same coin, but separate, independently testable capabilities. For a latent molecular rollout to serve as a causal world model, it must support both: knowing which structural direction produces a desired property change, and how large a displacement is needed to produce a calibrated magnitude of that change. Reporting a single cosine similarity or scalar MAE is therefore insufficient to claim a ``controllable'' latent space. Directional success does not guarantee calibrated property changes, and conclusions about amplitude can reverse entirely depending on which readout function is used.

The readout dependence we observe is not a flaw of this specific architecture---it is a general property of high-dimensional representations. Any scalar readout defines a single direction in latent space; if that direction does not align with the axes along which molecules actually vary, it will fail to capture amplitude even when the policy direction is correct. The frozen ridge null result is a clear demonstration: a readout can have excellent scalar prediction accuracy while being geometrically blind to the displacements that differentiate molecules. Its near-zero amplitude correlation is mechanically forced, not evidence that the policy direction carries no property signal.

This pattern extends beyond chemistry. Activation steering and representation engineering locate a direction with a probe, apply it, and validate the intervention by a scalar effect; our readout-family result suggests a concrete check that is usually omitted. If the probe gradient has negligible projection onto the displacement set being controlled, the reported steering strength measures the probe's geometry rather than the representation. Because alignment of an independently fitted readout concentrates at $D^{-1/2}$ by default, such nulls should be expected and reported rather than attributed to the model's capacity. The diagnostic is cheap: fit a family of readouts on frozen activations, and compare gradient--displacement alignment against amplitude recovery before claiming calibrated control.

We emphasize that the nonlinear \texttt{PkaHead} result recovers only relative amplitude ordering, not physically calibrated magnitude. The compressed scale (median predicted $2.33$ vs.\ true $3.03$) and the per-pair scaling error mean that we cannot claim true quantitative amplitude control---only that the policy's step length and direction jointly track the relative size of the requested property change. Absolute calibration remains an unsolved problem.

The text shortcut (\cref{sec:data_shortcut}) and the readout dependence of the amplitude result are two instances of the same readout--explanation gap: a successful scalar readout need not establish either use of the intended input pathway or recovery of the intended physical quantity.

\subsection{Metric structure is not operation-level semantics}

The measured geometry supports an auditable \emph{metric structure}: interpolation is smooth, multi-step optimization reduces error, the policy direction is above chance, and PCA places most displacement variance inside the high-variance subspace. These are local geometric facts under the tested readouts.

The stronger question is whether a latent direction corresponds to an identifiable chemical operation. That correspondence is not established here: naive gradients are weak, the exploratory shared-direction estimate has an interval crossing zero, and the amplitude estimate changes with the readout. We therefore claim metric structure, not pre-validated operation semantics. A future external yardstick, such as Hammett $\sigma$ or another linear free-energy relationship \citep{Hammett1937}, could test whether signed property changes follow an independently fitted substituent relationship on held-out chemical families. Unlike simple additive benchmarks, such a test would verify that the learned direction corresponds to a consistent physical mechanism, not just a data-driven correlation.

\subsection{What remains for full causal rollout?}

A complete rollout would require joint direction--amplitude success on independently selected pairs, reliable decoded structural changes, and chemical validation. The present work reaches the direction bar but not a calibrated amplitude bar. At the decoding tier, our audits of input-slot editing and final-state injection establish clear bounds on the tested control surfaces: even near-perfect alignment of the pooled latent state is insufficient to drive structural changes at the evaluated perturbation magnitudes. This does not imply a fundamental decoder-wide bottleneck---alternative intervention points such as ChemToken-level editing or persistent key--value cache perturbations may yet yield structural control. It does, however, demonstrate that latent-space controllability does not automatically transfer to output structures, and that decoding-side transfer must be validated explicitly rather than taken for granted.

Generation has a parallel boundary, and it is narrower than the direction result. One-epoch SFT improves over gen2-final under learned-oracle evaluation, while its advantage over the train-selected constant stays below the resolution of that test at this sample size. DPO reduces copying and approaches, without reaching, its direction design reference, and extending the budget does not close that gap. Best-of-8 remains an oracle upper bound rather than deployable control. Each statement is about one recipe under one evaluation; together they delimit the tested recipes rather than the space of preference objectives, data constructions, or selection rules, and they single out the quantity a next iteration has to move. The same boundary governs the external validation loop comparison: whether an internal audit reduces oracle calls is untested, and our two control-surface nulls indicate that a naive latent-to-structure path would not substitute for external validation on its own.

\subsection{Implications for molecular rollout}

Direction and amplitude must be designed, optimized, and reported as separate axes. Scalar fit, naive gradients, PCA, and explicit policy supervision each answer a different question. The central finding is that directional alignment is measurable, that it survives a pre-registered expansion of the evaluation bank, and that amplitude estimates depend on the readout, with no calibrated scale established so far. The audit itself is reusable: the pair-bank rule is written down and extends without refitting, and the readout family costs minutes on frozen activations, so the same three tiers can be applied to other latent molecular models. Future work should prioritize amplitude-aware training and selection, reliable decoded-state control, external chemical validation, non-oracle generation control, and independently selected pairs rather than additional variations of the same gradient diagnostic.

\section{Limitations}
\label{sec:limitations}

\paragraph{Split and external validity.} The primary train/test split shares 247 non-empty scaffolds and is not strictly scaffold independent; ridge fitting removes train--test scaffold overlap, but the frozen backbone was trained before this readout-specific cleaning. Policy validation is scaffold grouped, yet the 43 test pairs informed earlier diagnostics and are not an untouched external cohort. A pre-registered 151-pair expansion over the same held-out states reproduces the direction result with a narrower interval, but it shares those states and is a power check rather than a replication on new molecules. Pair-bootstrap intervals do not model all dependence from shared molecules or cross-fitted directions, and no wet-lab or independent prospective replication is included. All findings are limited to iBonD $\pka$, this LatentChem architecture, and the evaluated checkpoints. Only aqueous-phase measurements are used; the DMSO and other non-aqueous iBonD records are excluded, so cross-solvent generalization is untested rather than negative.

\paragraph{Readout and control-surface scope.} Direction comes from an explicitly direction-supervised policy, not an emergent signal. Amplitude is readout-dependent: the selected frozen ridge is nearly orthogonal to displacements ($1.18\times10^{-5}$), so its correlation $-0.0199$ is mechanically likely; the nonlinear \texttt{PkaHead} reread gives $0.460$ at a unit step and $0.585$ for the learned norm, with compressed scale ($2.33$ vs $3.03$ true) and no recovery for the true displacement direction ($\rho=-0.093$). A single global factor fitted on the 5{,}456 training pairs does not calibrate held-out pairs, so the residual is per-pair scale error. P3 (input-slot editing) and direct final-state injection both fail at the tested control surfaces and sizes; ChemToken-level editing and other surfaces remain untested. Ridge and PCA are sensitivity analyses, and G4's cosine bands are descriptive conventions rather than universal thresholds. We also do not evaluate oracle-cost reduction: no external validation loop baseline is run, and the practical claim that an internal audit could substitute for or reduce external calls remains untested.

\paragraph{Generation and selection scope.} Generated structures and candidate selection are scored with the repaired learned $\pka$ oracle; generated molecules are not experimentally assayed. SFT and DPO constant comparisons use the train-selected constant $3.74$ over all 255 rows with missing predictions treated as no-op, while the SFT--gen2-final paired test uses 235 both-scored rows. Best-of-8 uses oracle target information and is not deployable. DPO uses $\beta=0.1$ with final-checkpoint evaluation at two and four epochs; these null comparisons do not exclude other preference data, optimization, or selection rules. Copy-rate changes are observed frequencies, not proof of a conservatism mechanism.

\section{Conclusion}
\label{sec:conclusion}

We asked how far a latent molecular rollout can move toward a causal world model, and the three-tier audit gives a specific answer: further than scalar accuracy suggests, less far than controllability requires. \textbf{Tier 1}---the repaired latent-inference path improves scalar prediction over direct molecular readout after controlling for random-position effects, so the rollout genuinely predicts through the molecule, though this does not by itself establish molecule-specific control. \textbf{Tier 2}---a target-conditioned policy reliably learns displacement direction, and the estimate survives a pre-registered $3.5$-fold expansion of the evaluation bank, whereas the amplitude interpretation changes with the readout: the frozen ridge readout is nearly orthogonal to the true displacements and cannot test amplitude, whereas the nonlinear \texttt{PkaHead} readout recovers relative amplitude order but not a calibrated magnitude. \textbf{Tier 3}---matching a target latent state does not guarantee a decoded structural change, so latent control does not yet survive the trip to structure. Direction is the part of causal control this rollout already has; amplitude and decoded transfer are the parts still missing.

The audit does more than score the system, because two of its boundaries come with a mechanism attached. Amplitude is decided by readout geometry rather than by fit quality: nine independently fitted readouts sit at $10^{-5}$ alignment and none recovers amplitude, whereas the pathway-co-trained head reaches $0.107$ and does. Decoded transfer fails at the tested interfaces while the pooled latent state itself is realized to cosine $0.99997$, which places the remaining difficulty in the latent-to-structure interface rather than in the displacement. What the next iteration needs is therefore specific rather than a list of retries: a calibrated joint direction--amplitude objective, an interface that carries latent edits into decoding, non-oracle generation control, where the current gains remain oracle-dependent, and chemical validation on independently selected pairs.

\section*{Reproducibility and drafting disclosure}

The accompanying \texttt{data/evaluation\_artifact\_manifest.json} records SHA-256 hashes for the linked shortcut-diagnostic, latent, readout, PCA, policy, P3-decoding, SFT, offline-DPO, figure, table, log, and analysis-code artifacts. Latent-dependent G4, shared-projection, ridge, and PCA calculations assert the same test-latent hash before combining values; SFT and DPO provenance instead records generation inputs and training artifacts because they do not read those latent files. P3 provenance records both decoding protocols, paired learned-oracle summaries, the explicit cohort-overlap analysis, and the slot-identity check on the initial pre-forward prompt embeddings. The direct-injection archive additionally records the hook-preflight check, every realized relative hidden-state displacement, and paired outputs for all four steps and both protocols. An AI assistant supported LaTeX assembly and language drafting; quantitative statements were checked against the accompanying experimental artifacts. The author is responsible for the final scientific interpretation and for any venue-specific disclosure requirements.

\begingroup
\small
\setlength{\bibsep}{3pt plus 1pt}
\bibliographystyle{plainnat}
\bibliography{references}
\endgroup

\clearpage
\appendix
\renewcommand{\thefigure}{S\arabic{figure}}
\renewcommand{\thetable}{S\arabic{table}}
\setcounter{figure}{0}
\setcounter{table}{0}

\section{Supplementary controls and diagnostics}
\setcounter{subsection}{-1}

\begin{figure}[h]
\centering
\includegraphics[width=\textwidth]{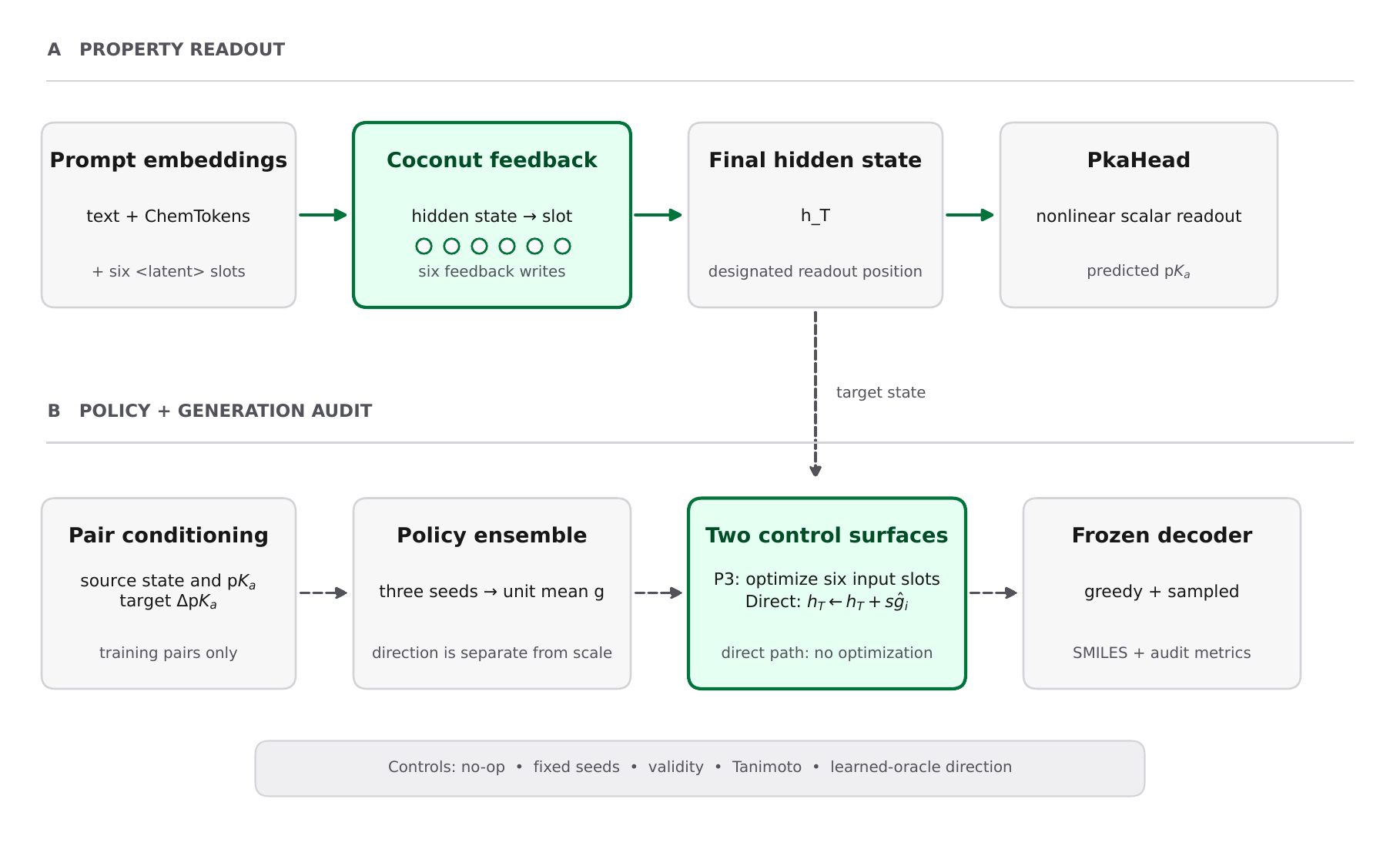}
\caption{Algorithm structure and control surfaces. (A) The property pathway combines text and ChemTokens with six latent input slots, performs six feedback writes, and reads a final hidden state with \texttt{PkaHead}. (B) The policy ensemble predicts a direction from pair conditioning. P3 (input-slot editing) optimizes the six input-slot embeddings to match a pooled target, whereas direct injection adds the policy direction to the final hidden state without optimization (\cref{sec:hidden_injection_methods}). Both paths are evaluated by the same frozen decoder under greedy and fixed-seed sampled decoding.}
\label{fig:algorithm_overview}
\end{figure}

\subsection{Moved methods and scope details}

\paragraph{Training-budget ablation and warm-start exclusion.} In the confirmatory budget ablation, the declared epoch count and corresponding cosine-schedule horizon are the only changes; data, initialization, initial learning rate, DPO $\beta$, seeds, and decoding protocols are fixed. The longer budget is primary only if it changes the qualitative conclusion. Longer-budget runs restart from the original initialization rather than continuing from the shorter-budget checkpoint: the cosine horizon is part of the recipe, saved checkpoints contain adapter and head weights but no optimizer or scheduler state, and per-epoch sampler seeding changes data order, so a warm start would confound the budget change with a learning-rate restart.
\label{sec:training_budget_appendix}

\paragraph{G4 band definitions and pair-bank distinction.} For $4096$-dimensional random unit directions, the cosine standard deviation is approximately $1/\sqrt{4096}\approx0.0156$, with a 95\% reference interval of approximately $[-0.031,0.031]$. The bands are descriptive conventions, not pass/fail thresholds: $\lvert c_i\rvert<0.05$ (random-direction magnitude), $0.05\leq c_i<0.2$ (weak positive direction), and $c_i\geq0.2$ (strong positive direction). Because pairs are oriented from lower to higher $\pka$, $c_i\leq-0.05$ indicates anti-alignment. As an upper-range check, no original top-20 pair exceeds $0.5$. The original 20-pair bank and expanded 43-pair bank are distinct selections and are not interchangeable estimates. Molecular pairs use Morgan-radius-2 2{,}048-bit similarity \citep{Rogers2010Extended}.
\label{sec:g4_bands_appendix}

\paragraph{Held-out pair-bank expansion, pre-registered.} The 43-pair bank is small enough that a median over it carries a wide interval, so we fixed an expansion rule before scoring any policy output on the larger bank. The original rule keeps every within-test pair with $|\Delta\pka|\geq1$ and Morgan radius-2 $2{,}048$-bit Tanimoto similarity at least $0.6$, oriented from lower to higher $\pka$; this is exactly the 43-pair bank used in Tier~2. The expanded rule changes one parameter, lowering the similarity floor to $0.5$, which is the floor already used when the training pair bank was built. Because the rule is otherwise identical and the threshold is relaxed, the expanded bank is a strict superset: $151$ pairs, comprising the original $43$ plus $108$ new pairs in the band $0.5\leq\mathrm{Tanimoto}<0.6$. Training pairs, policy architecture, seeds, epoch budget, and checkpoint selection are untouched; the trained policy is only re-scored, so no readout or policy parameter is fitted on the new pairs. Results are reported separately for the original $43$, the new $108$, and the pooled $151$ so that any threshold dependence remains visible, and the original $43$ remain the primary estimate. This expansion raises statistical power over the same $1{,}151$ archived held-out states; it does not create an independent external cohort.

\paragraph{Amplitude-reread decision rule.} The decision rule was fixed before execution: a \texttt{PkaHead} correlation above $0.5$ would overturn the amplitude conclusion, values near zero under both readouts would confirm it, and disagreement would downgrade the claim to readout dependence.
\label{sec:amplitude_reread_appendix}

\paragraph{P3 diagnostics and checkpoint scope.} For the paired learned-oracle analysis, source, baseline, and intervened SMILES are scored by the archived MAE-1.1 text-query PkaHead. The intervention effect is $\Delta\hat y_i=\hat y_i^{\mathrm{int}}-\hat y_i^{\mathrm{base}}$; unchanged outputs therefore have zero effect and count against directional success. If an identical output pair cannot be oracle-scored, its zero effect follows from output identity rather than an imputed numeric prediction. The fixed checks are validity $\geq0.9$, median source-anchored Tanimoto $\geq0.6$, sign accuracy $\geq0.65$, and median $|\Delta\hat y|\in[0.2,1.0]$. The property and generator checkpoints have different \texttt{PkaHead} and projector tensors (maximum absolute difference $0.246$ in \texttt{PkaHead}), so P3 tests cross-checkpoint transfer. The pooled target does not uniquely determine the six slot embeddings. A separate identity check reads $z_i=\texttt{pe}[0,\texttt{latent\_pos}]$ immediately after \texttt{get\_prompt\_embeddings}, before any backbone forward and before Adam optimization. Across 20 held-out sources and 190 comparisons, each $z_i$ is a $6\times4096$ input-embedding block with median cosine $0.99978$ (minimum $0.99956$) and median relative $L_2$ distance $0.021$ (maximum $0.030$). These are molecule-invariant initial input embeddings, not optimized slots or feedback-written hidden states. The pass-count diagnostic separately observes seven backbone calls from coconut feedback; P3 decoding calls the generator directly with \texttt{inputs\_embeds} and does not execute that feedback loop.
\label{sec:p3_diagnostics_appendix}

\paragraph{Hidden-injection preflight.} A preflight example verifies that the hooked vector matches the pooled $h_T$ (cosine $1.0$, relative error $0$) and that a step of $0.1$ changes the first-token logits ($L_2$ difference $1.854$), although the top-1 token remains unchanged.
\label{sec:hidden_preflight_appendix}

\paragraph{Cross-fitted shared-direction projection.} Let $u_i=d_i/\lVert d_i\rVert_2$ denote each low-to-high unit displacement. For held-out pair $i$, let $\mathcal S_i$ contain every other pair that shares a non-empty scaffold with either endpoint of $i$, and define $\mathcal T_i=\{j\neq i:j\notin\mathcal S_i\}$. The leave-scaffold-out estimate and projection are $\bar u_{-i}=(\sum_{j\in\mathcal T_i}u_j)/\lVert\sum_{j\in\mathcal T_i}u_j\rVert_2$ and $p_i=u_i^\top\bar u_{-i}$. The sign-flip null independently reverses training-pair orientations within each fold, recomputes $\bar u_{-i}$ and $p_i$, and tests the observed mean projection one-sided. Ordinary leave-one-pair-out analysis is a sensitivity check. This is cross-fitted evidence over the same 43 test pairs, not independent train--test validation.
\label{sec:shared_projection_appendix}

\paragraph{Ridge protocol.} From the original training split, rows sharing a non-empty Murcko scaffold or exact canonical SMILES with test are removed. The remaining 3{,}369 rows are divided by scaffold into 2{,}855 fit and 514 validation rows using GroupShuffleSplit (test size $0.15$, seed 42); empty scaffolds are grouped by exact canonical SMILES. The original backbone was trained before this readout-specific cleaning, so the complete model is not strictly scaffold isolated. If $s$ is the fit-standard-deviation vector, $z=(h-\mu)/s$, $w$ is the original-$h$ readout gradient, and $d=h_{\mathrm{high}}-h_{\mathrm{low}}$, then $w_z=w\odot s$, $d_z=d\oslash s$, and $w_z^\top d_z=w^\top d$. Scalar prediction is unchanged, but cosine is not invariant under dimension-wise standardization.
\label{sec:ridge_protocol_appendix}

\paragraph{Readout-family and displacement-cone protocol.}
\label{sec:readout_family_methods}
All nine readouts are fitted on the same standardised latents ($9{,}202$ train rows, per-dimension mean and standard deviation from train) with validation-MAE-only selection, and none sees test labels. Ridge heads use penalties $\alpha\in\{0.01,0.1,1,10,100\}$; the two neural heads are one-hidden-layer ReLU regressors trained with Adam, batch $128$, at most $120$ epochs, and early stopping on validation MAE with patience $10$. Controls are a ridge fitted on permuted labels and a ridge fitted on a random half of the training rows (seed $0$). Alignment is the mean cosine between the readout gradient at the source state and the unit displacement $d/\lVert d\rVert$ over the 43 held-out pairs used throughout Tier~2; amplitude is $\Delta=f(h_b)-f(h_a)$ correlated (Spearman) with the true $\Delta\pka$. Neural gradients are exact input gradients from backpropagation, not finite differences. The cone statistics use the same 43 unit displacements: pairwise cosines, a singular-value participation ratio, the number of leading dimensions reaching $90\%$ of variance, and the angle between the selected ridge gradient and the mean displacement axis. The weight-energy check projects the selected ridge gradient onto the six leading principal components (the $P_{90}$ subspace of the PCA test) and onto an orthonormal basis of the unit displacements, and reports both against the expected energy fraction $k/D$ of a random direction in a $k$-dimensional subspace; it is repeated after deleting the two coordinates that carry $99.99\%$ of the ridge energy.

\paragraph{PCA protocol.} PCA is fit on raw $h$ using only the 2{,}855 fit rows, without per-dimension scaling or access to test labels. $P_{90}$ projects onto the smallest leading subspace explaining at least 90\% of fit variance. Test pairs do not construct the mean-fit direction. Subspace coverage and directional agreement are therefore not conflated.
\label{sec:pca_protocol_appendix}

\begin{table}[h]
\centering
\caption{Evaluation cohort scope. Per-experiment $n$ values appear here rather than repeated across the main text.}
\label{tab:scope}
\small
\begin{tabular}{@{}p{0.24\linewidth}p{0.28\linewidth}p{0.42\linewidth}@{}}
\toprule
Analysis & Cohort sizes & Role \\
\midrule
Five-group regression & 9{,}202 train; 1{,}151 test & Scalar contribution to the rollout system \\
G1/G2/G3/G5 & 38 pairs; 30 trajectories; 40 samples; 105 pairs & Geometry, intervention, and directional-sign evidence \\
G4 pair banks & Original 20 pairs; expanded 43 pairs & Separate selections; not interchangeable estimates \\
Ridge/PCA/shared projection & 2{,}855 fit; 514 validation; 381 fit and 43 test pairs & Frozen readout, subspace, and shared-direction tests \\
Policy head & 5{,}456 training pairs; 43 held-out pairs & Direction, fixed-step amplitude, and amplitude supervision \\
P3 / direct injection & 80 sources & Cross-checkpoint control-surface audits \\
Generation & 2{,}295 SFT rows; 255 eval rows; 235 paired rows & Paired generation improvement and oracle headroom \\
\bottomrule
\end{tabular}
\end{table}

\subsection{Text-shortcut diagnosis (pre-repair)}

\begin{figure}[h]
\centering
\includegraphics[width=0.82\textwidth]{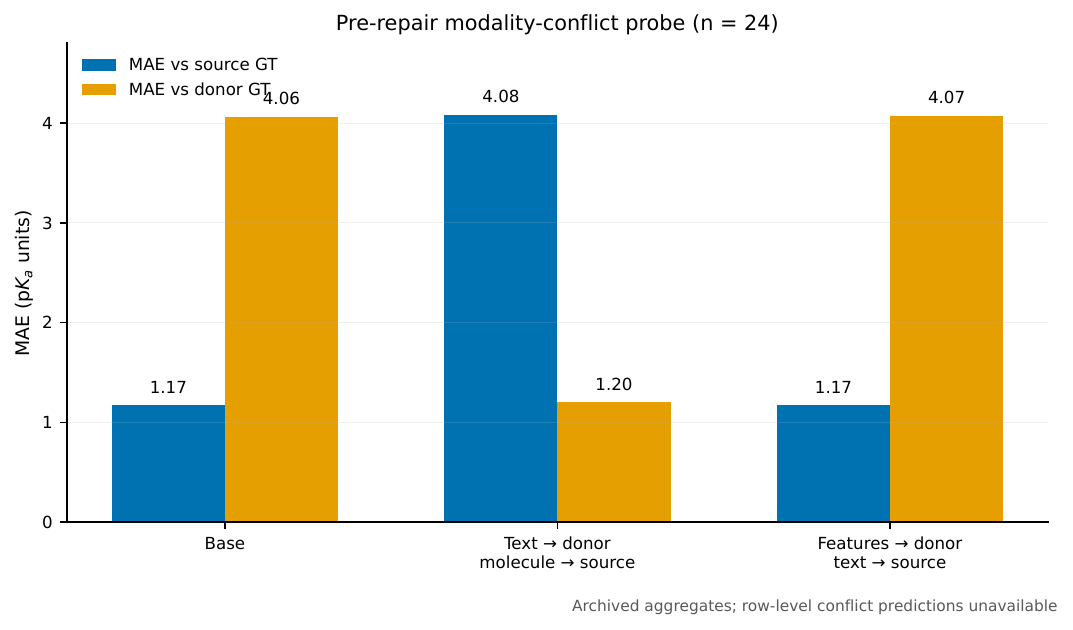}
\caption{Archived modality-conflict summary for the pre-repair C configuration ($n=24$). Text replacement reverses source- and donor-referenced error, whereas molecular-feature replacement leaves source MAE at $1.17$. Only aggregate MAEs were preserved; per-row conflict predictions are unavailable, so no intervals or invariance counts are inferred.}
\label{fig:shortcut}
\end{figure}

The pre-repair evidence comprises two distinct records. The archived \texttt{ref\_head} predictions in \texttt{fig3\_causal.json} give MAE $1.103$ on 24 held-out diagnostic rows; this is not a full-test-set score. Separately, \texttt{DIAGNOSIS.md} preserves aggregate source- and donor-referenced MAEs for the modality-conflict probe ($n=24$). Replacing the query SMILES moves the error toward the donor target, whereas replacing the SMI-TED features leaves source error unchanged at the reported precision. The conflict record does not contain per-row predictions, so \cref{fig:shortcut} has no uncertainty intervals and we report no output-invariance count.

\input{tables/tableS0_text_shortcut.tex}

\begin{figure}[p]
\centering
\includegraphics[width=0.98\textwidth]{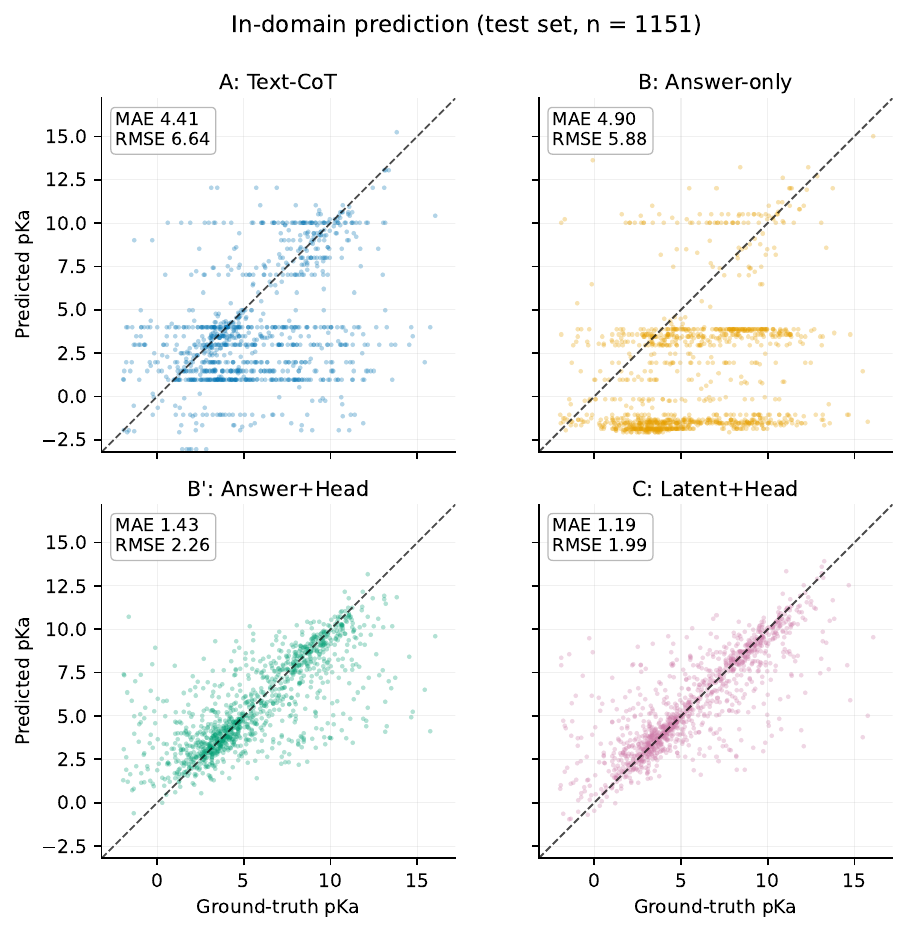}
\caption{Archived pre-repair in-domain prediction baseline on 1{,}151 rows. The embedded MAEs are A $4.414$, B $4.901$, B$'$ $1.434$, and C $1.191$, with respective RMSEs $5.918$, $6.381$, $1.837$, and $1.532$; all values are reproduced from \texttt{figdata.json}. This figure documents the scalar-prediction baseline from which the subsequent interpolation, rollout, and intervention diagnostics were developed. It is not the repaired five-group evaluation in \cref{fig:regression}: the checkpoint, prompt, and token alignment differ, so these legacy values are not used for the current architecture ranking.}
\label{fig:legacy_indomain}
\end{figure}

Two repairs were applied before the reported five-group comparison: SMILES removal from the query and a token-id alignment fix that restored recognition of all six latent slots rather than one. The pre-repair probe and repaired test numbers use different cohorts, and the two repairs were not separately ablated. Their numerical difference therefore cannot isolate the effect of shortcut removal; the conflict probe directly supports text dominance only in the archived pre-repair setting. For completeness, the project history records an evaluation-time truncation fix in the text-parsing path, affecting text-path scoring only; its raw diagnostic is not preserved, and it is not part of the \texttt{PkaHead}-based comparisons.

\subsection{Molecular-feature and latent-path controls}

\begin{figure}[h]
\centering
\includegraphics[width=0.9\textwidth]{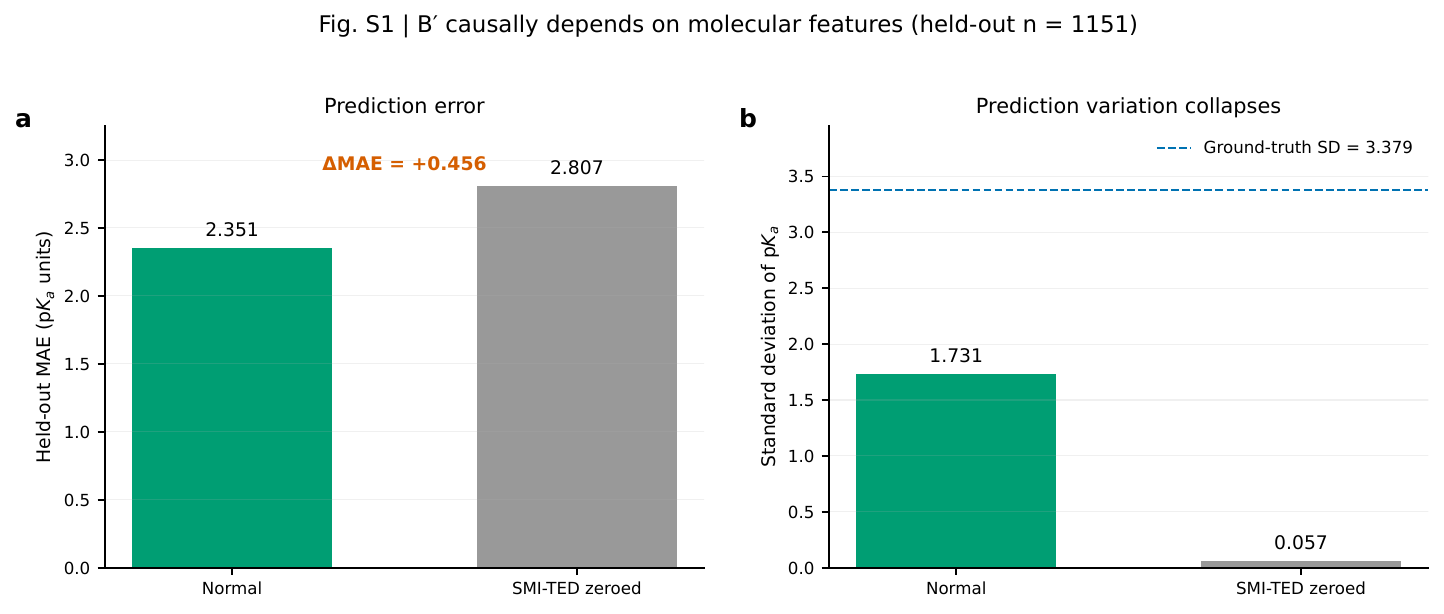}
\caption{Zeroing the SMI-TED molecular features in B$'$ degrades held-out prediction, supporting causal dependence of the repaired head on molecular input rather than text-only SMILES reading.}
\label{fig:s1}
\end{figure}

\begin{figure}[p]
\centering
\includegraphics[width=0.78\textwidth]{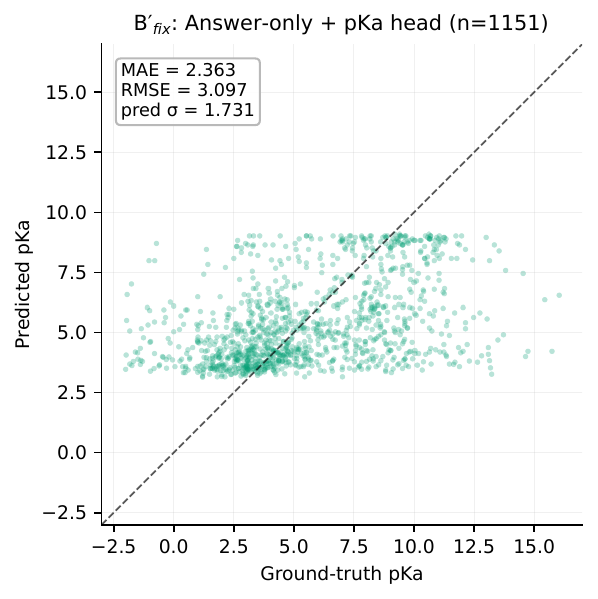}
\caption{Detailed repaired-B$'$ prediction scatter on the same 1{,}151-row test set as \cref{fig:regression}. Independent recomputation gives MAE $2.363$, RMSE $3.097$, and prediction standard deviation $1.731$, matching the embedded values. This full-resolution view complements the five-model comparison and documents the compressed prediction range against which the later latent-intervention metrics are interpreted.}
\label{fig:bp_fix_scatter}
\end{figure}

\begin{figure}[h]
\centering
\includegraphics[width=\textwidth]{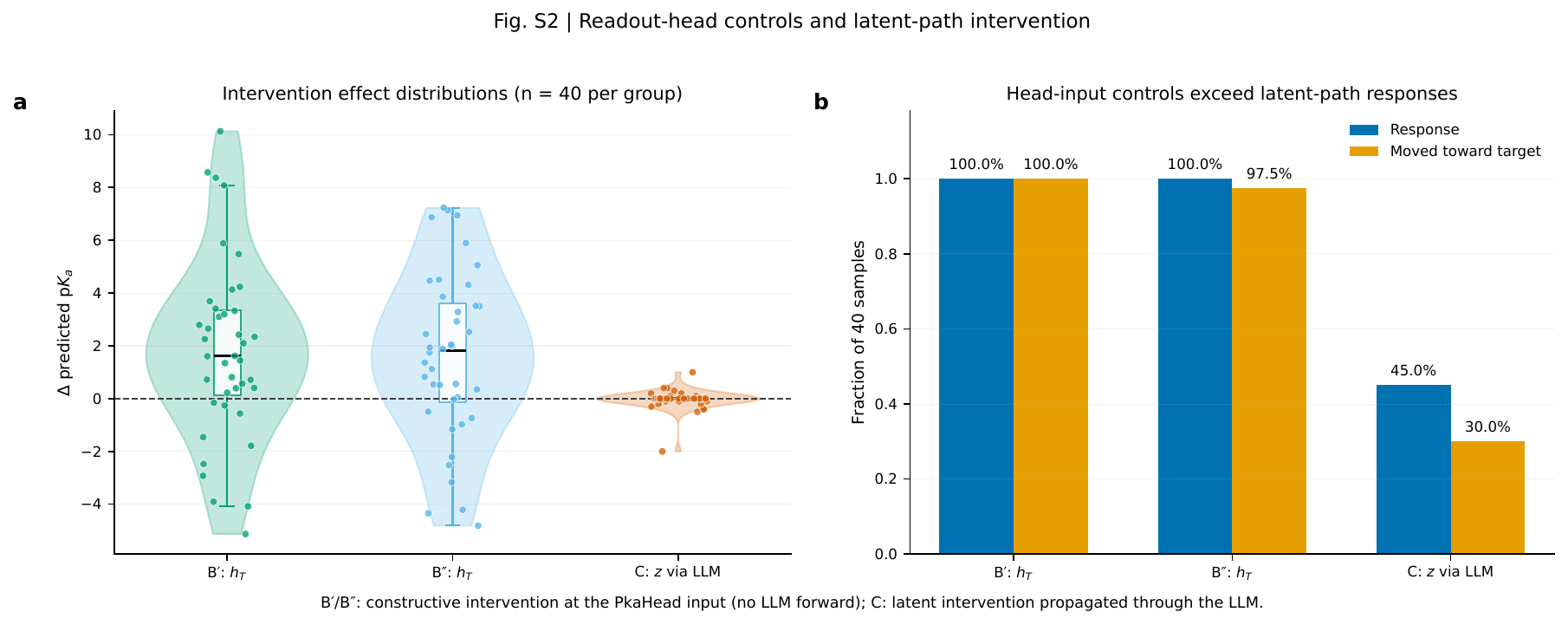}
\caption{Head-input intervention controls for B$'$ and B$''$ compared with the C-group latent intervention. Direct head-input perturbations can move scalar predictions without establishing directional molecular generation.}
\label{fig:s2}
\end{figure}

\begin{figure}[h]
\centering
\includegraphics[width=\textwidth]{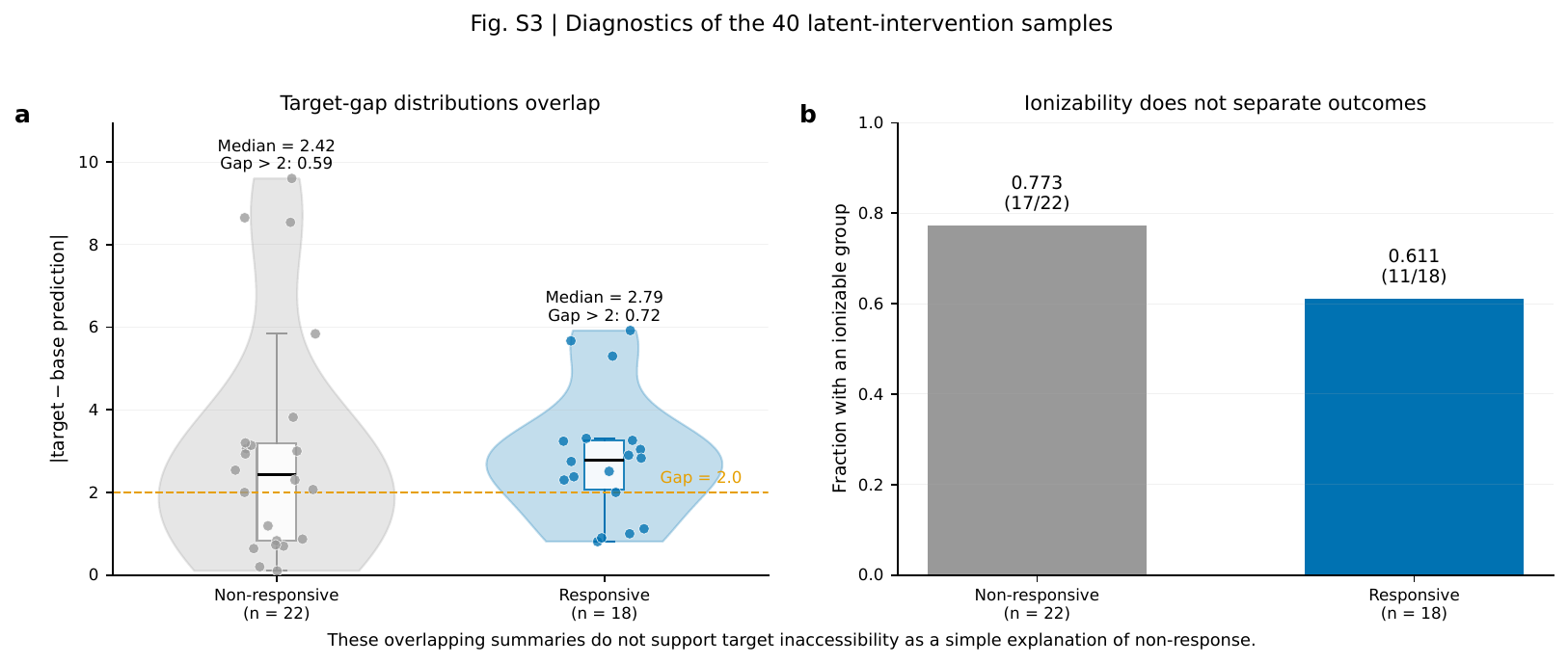}
\caption{Diagnostics of responsive and non-responsive samples in the 40 C-group interventions. The panels examine target gaps, ionizable groups, and response behavior underlying \cref{tab:nonresponse}.}
\label{fig:s3}
\end{figure}

\begin{figure}[h]
\centering
\includegraphics[width=\textwidth]{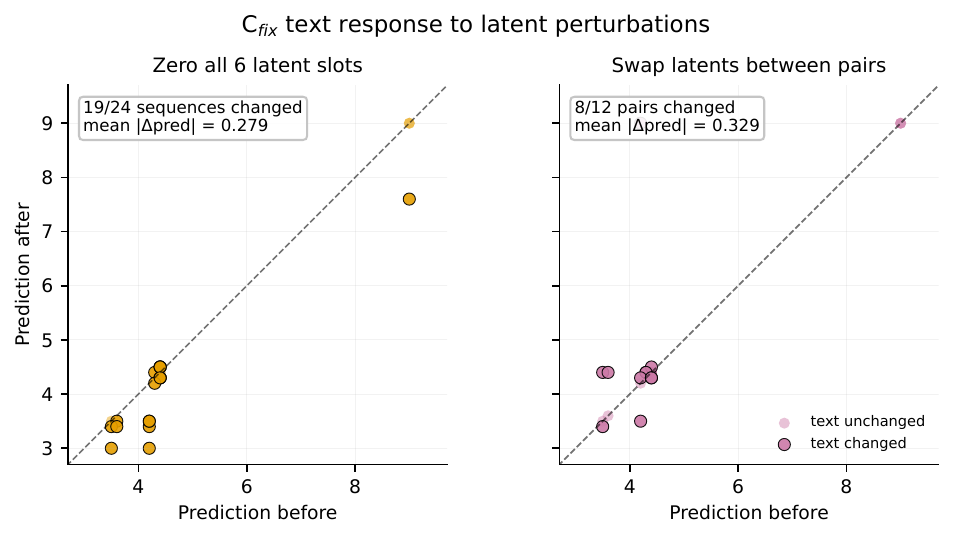}
\caption{Independent small-sample token-repair controls, distinct from the main 40-sample G3 cohort. Zeroing all six latent slots changes 19/24 generated sequences, with mean $\lvert\Delta\mathrm{prediction}\rvert=0.279$; swapping latent states changes at least one generation in 8/12 pairs. These controls support a functioning latent pathway after token repair while also showing incomplete response coverage.}
\label{fig:s4}
\end{figure}

Figures~\ref{fig:s1}--\ref{fig:s4} collect the controls for molecular-feature dependence, head-input intervention, non-responsive interventions, and latent zero/swap perturbations. These controls are small-sample mechanistic diagnostics and are not independent performance benchmarks.


\subsection{Data and training audits}

\begin{figure}[h]
\centering
\includegraphics[width=0.86\textwidth]{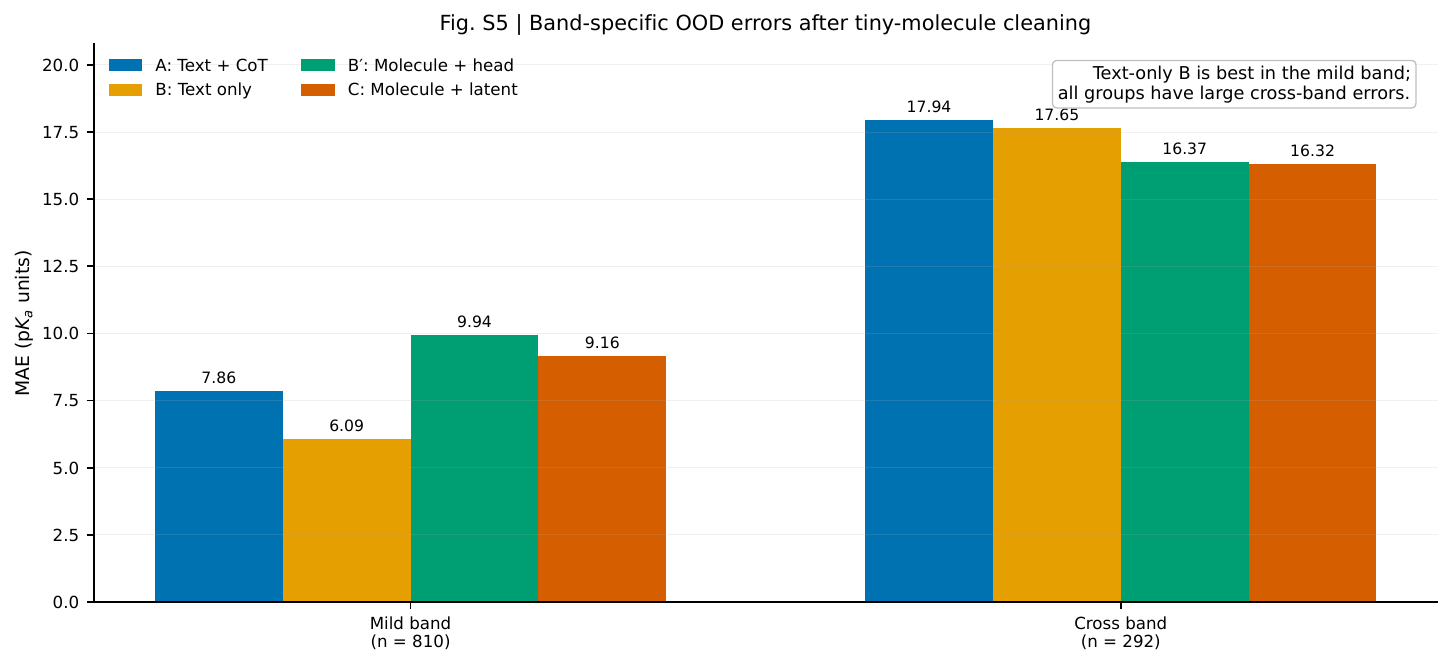}
\caption{Band-specific out-of-distribution errors after tiny-molecule cleaning. These strata describe error variation within the current data setting and are not a substitute for an independent scaffold-external test.}
\label{fig:s5}
\end{figure}
\begin{figure}[p]
\centering
\begin{subfigure}{0.48\textwidth}
\centering
\includegraphics[width=\linewidth]{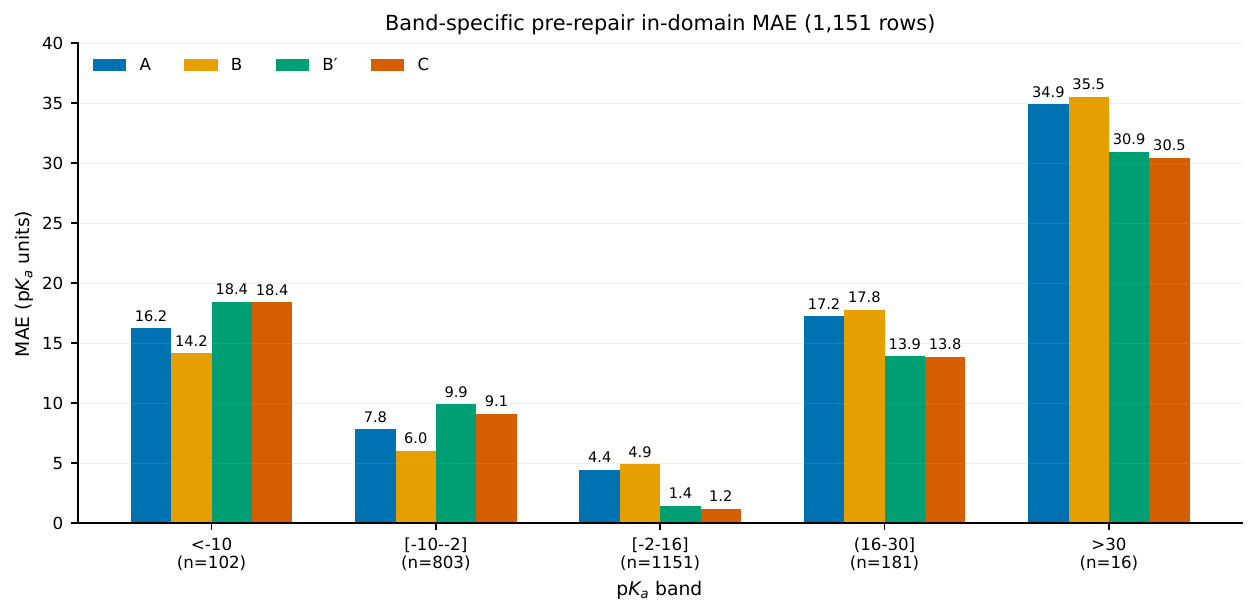}
\caption{Five-band scalar-prediction diagnostic. Values are recomputed from \texttt{figdata.json}; in-domain MAE is $4.4/4.9/1.4/1.2$ for A/B/B$'$/C. These archived pre-repair values are not pooled with the repaired in-domain comparison.}
\end{subfigure}\hfill
\begin{subfigure}{0.48\textwidth}
\centering
\includegraphics[width=\linewidth]{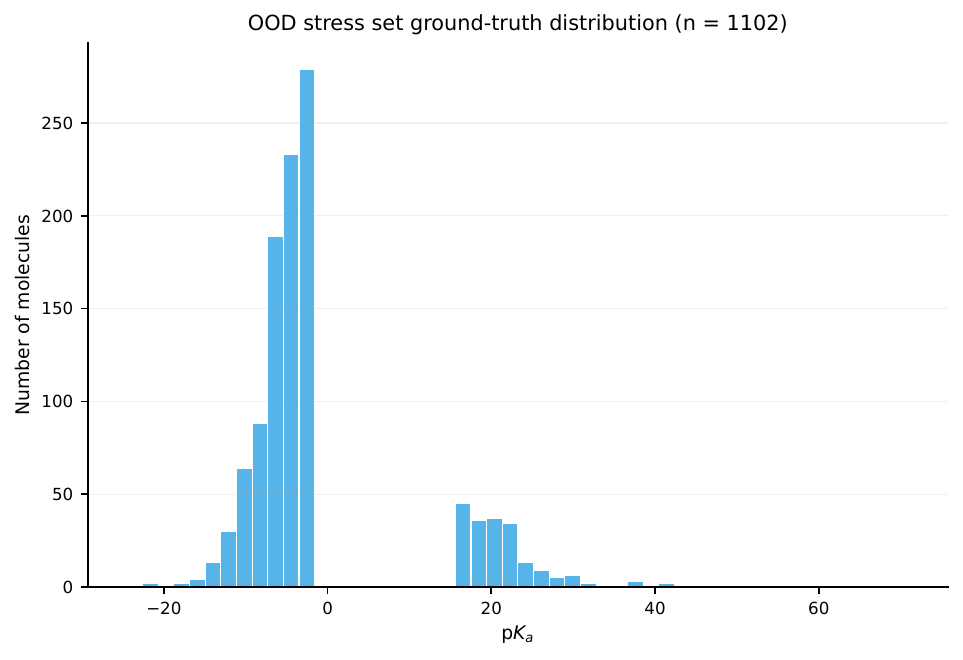}
\caption{Ground-truth $\pka$ distribution for the 1{,}135-row OOD stress set. $15.5\%$ of rows lie below $-10$ and $1.8\%$ above $30$; this is a composition diagnostic, not a balanced-coverage claim. Values outside roughly $[-2,16]$ are dominated by non-aqueous or computed entries and are shown only as composition diagnostics; they are not pooled with the aqueous in-domain comparison.}
\end{subfigure}
\caption{Out-of-domain error and dataset composition.}
\label{fig:ood_composition}
\end{figure}

\begin{figure}[p]
\centering
\begin{subfigure}{0.49\textwidth}
\centering
\includegraphics[width=\linewidth]{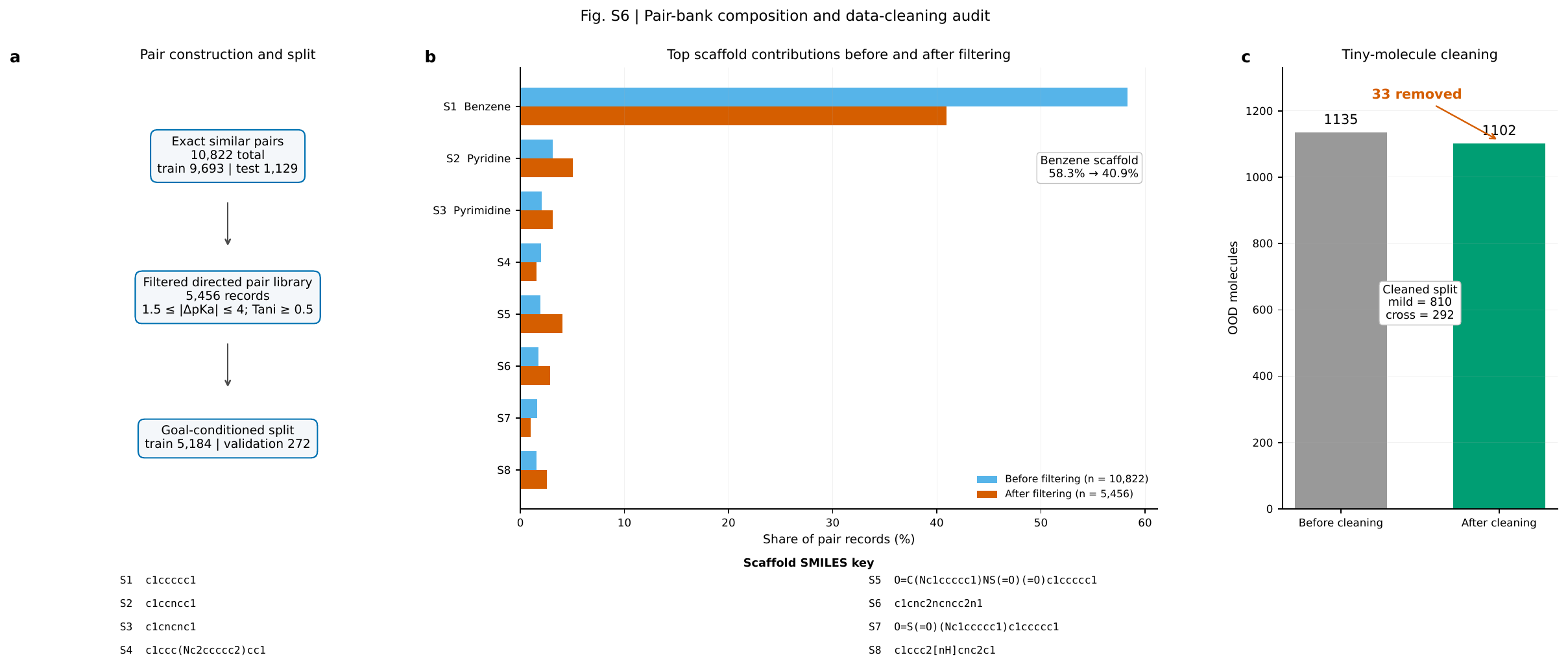}
\caption{Pair-bank composition and data-cleaning audit used by the latent-direction analyses.}
\end{subfigure}\hfill
\begin{subfigure}{0.49\textwidth}
\centering
\includegraphics[width=\linewidth]{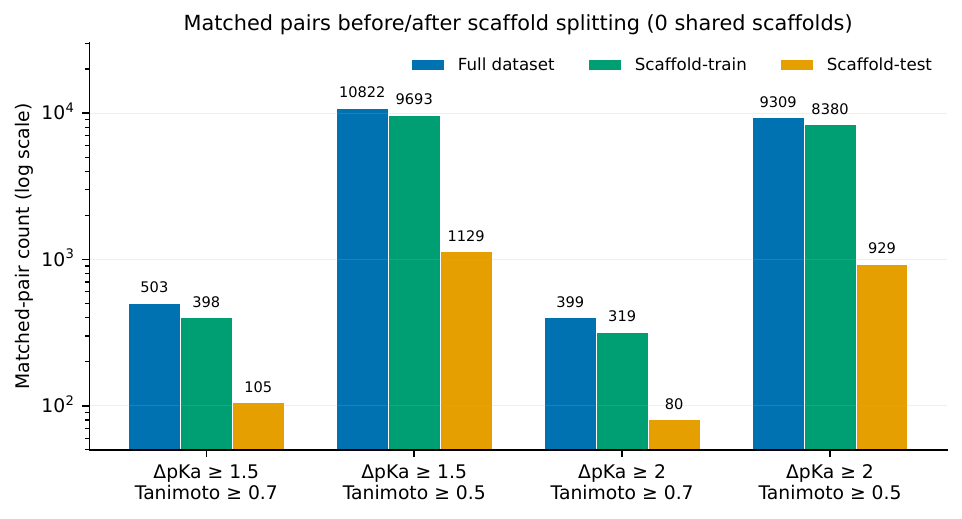}
\caption{Archived pair-library threshold and split visualization. The full-bank counts $10{,}822$, train $9{,}693$, and test $1{,}129$ agree with the current audit; the threshold panels are not the frozen 43-pair policy test set.}
\end{subfigure}
\caption{Pair-bank provenance and selection thresholds.}
\label{fig:pair_bank}
\end{figure}

\begin{figure}[h]
\centering
\includegraphics[width=0.86\textwidth]{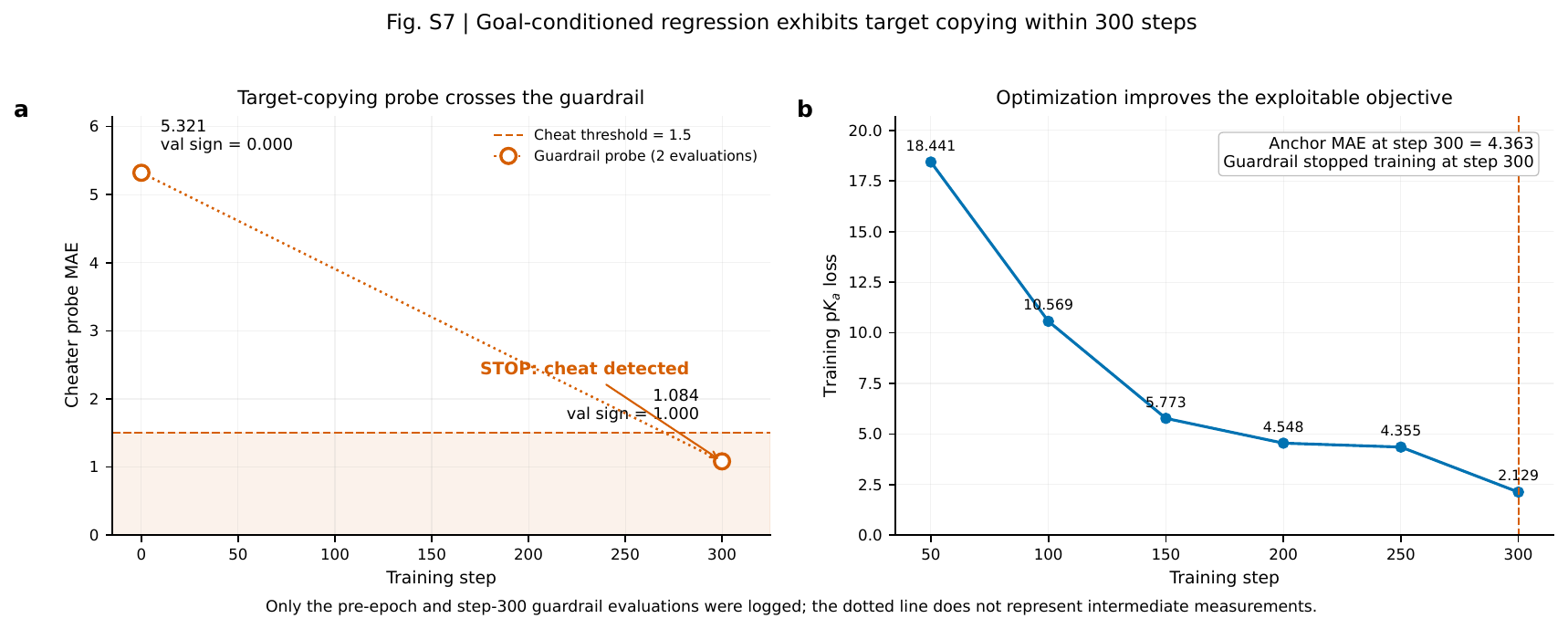}
\caption{Goal-conditioned regression guardrail: target copying emerges within 300 optimization steps in the diagnostic model. This control motivates explicit no-op, constant, and source-anchor checks in the generation analysis.}
\label{fig:s7}
\end{figure}

\clearpage
\section{Supplementary directional analyses}

\begin{figure}[h]
\centering
\includegraphics[width=\textwidth]{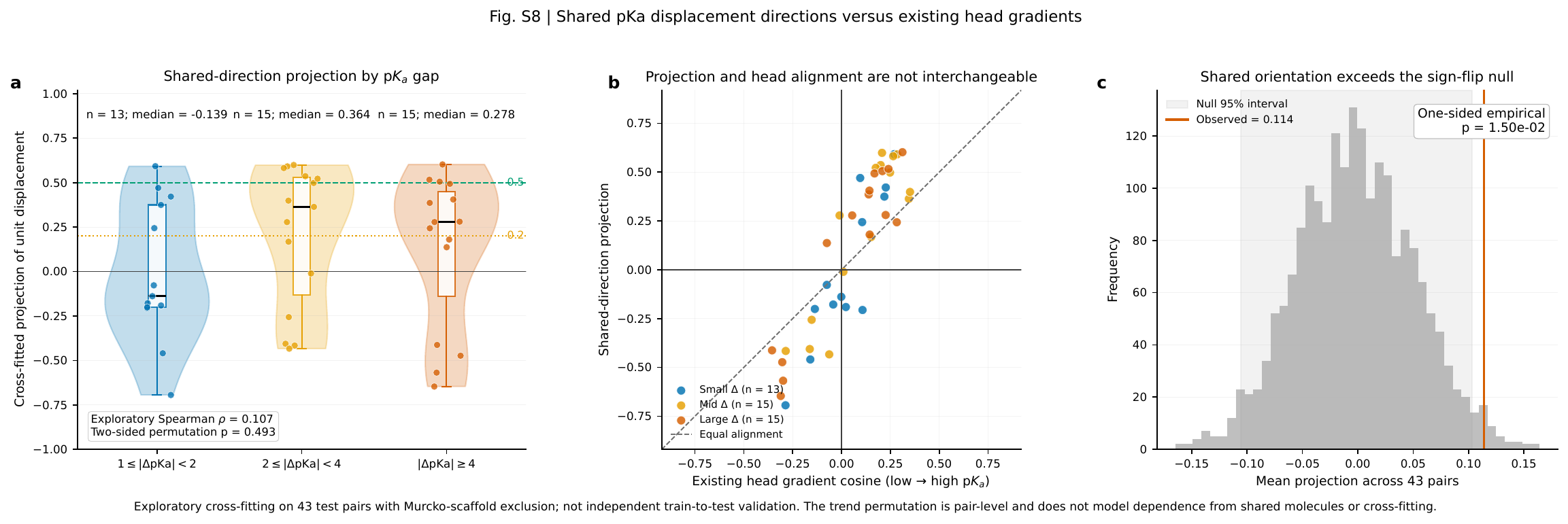}
\caption{Cross-fitted shared-direction projections compared with existing-head G4 cosines. The primary leave-scaffold-out median is $0.244$ (95\% CI $[-0.078,0.386]$). The one-sided sign-flip $p=0.0150$ tests the observed mean projection, not the median. This exploratory cross-fitting over known test pairs is not independent train--test validation.}
\label{fig:s8}
\end{figure}

\begin{figure}[h]
\centering
\includegraphics[width=\textwidth]{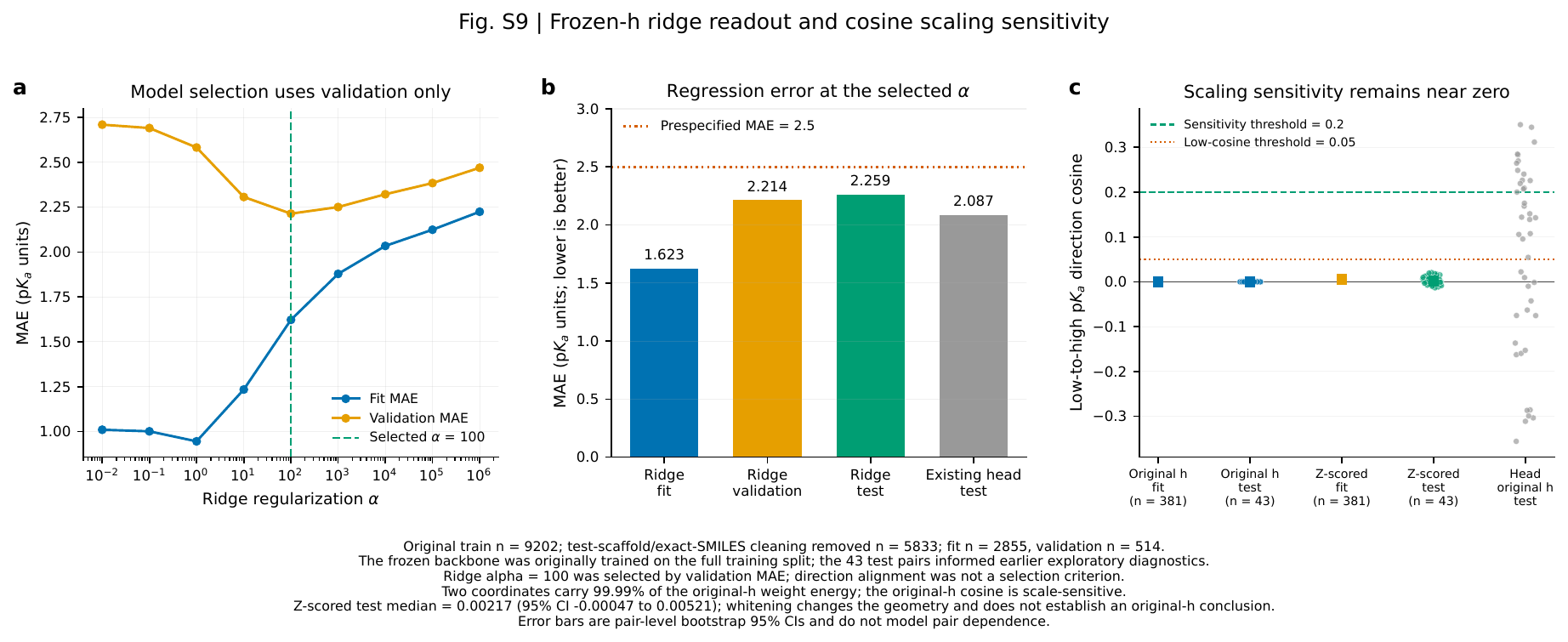}
\caption{Frozen ridge readout. (a) Alpha is selected by validation MAE only. (b) Scalar MAE at $\alpha=100$. (c) Original-$h$ and $z$-scored ridge cosines, with the original-$h$ existing-head values shown separately. The two ridge cosine geometries are not directly interchangeable; alpha selection did not use directional alignment.}
\label{fig:s9}
\end{figure}

\begin{figure}[h]
\centering
\includegraphics[width=\textwidth]{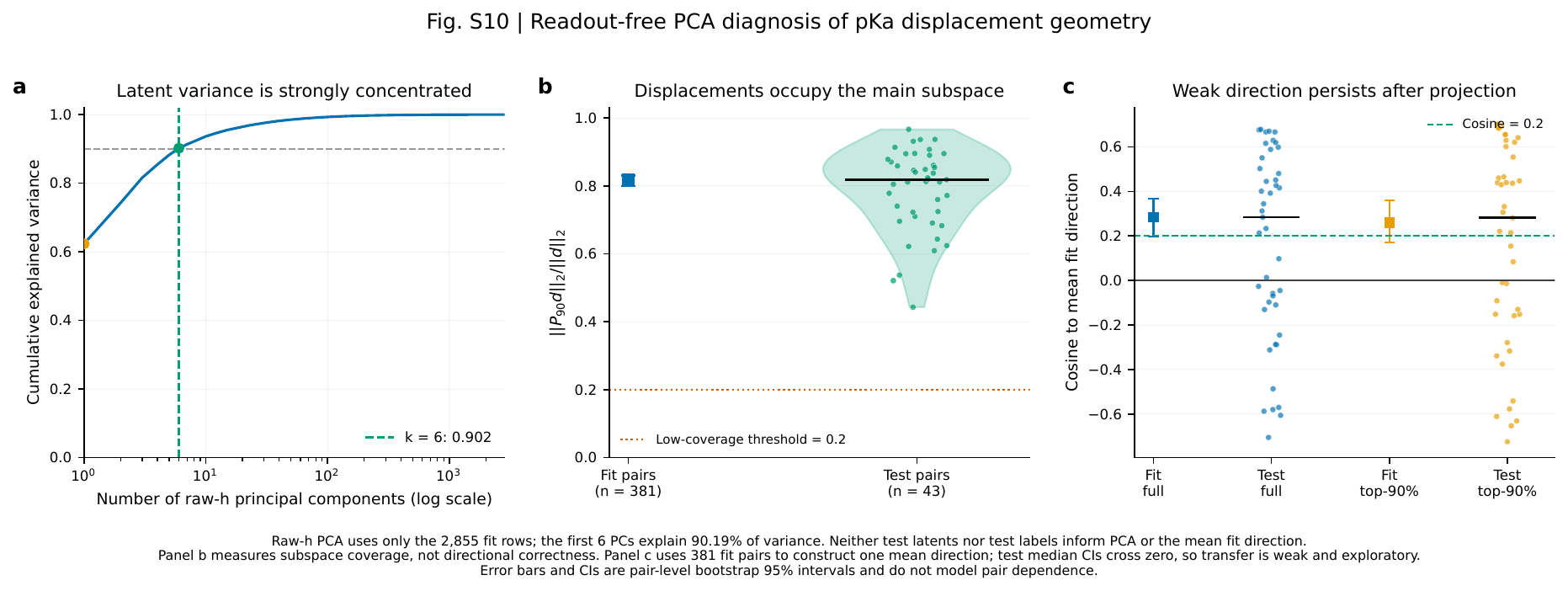}
\caption{Readout-free PCA direction analysis. (a) Six components explain $90.19\%$ of fit variance. (b) Median test-displacement norm retained in the top-90\% subspace is $0.818$; this is coverage, not direction accuracy. (c) Mean-fit-direction transfer has median cosine $0.284$ in full space and $0.282$ after projection, with both intervals crossing zero.}
\label{fig:s10}
\end{figure}

\begin{figure}[h]
\centering
\includegraphics[width=0.76\linewidth]{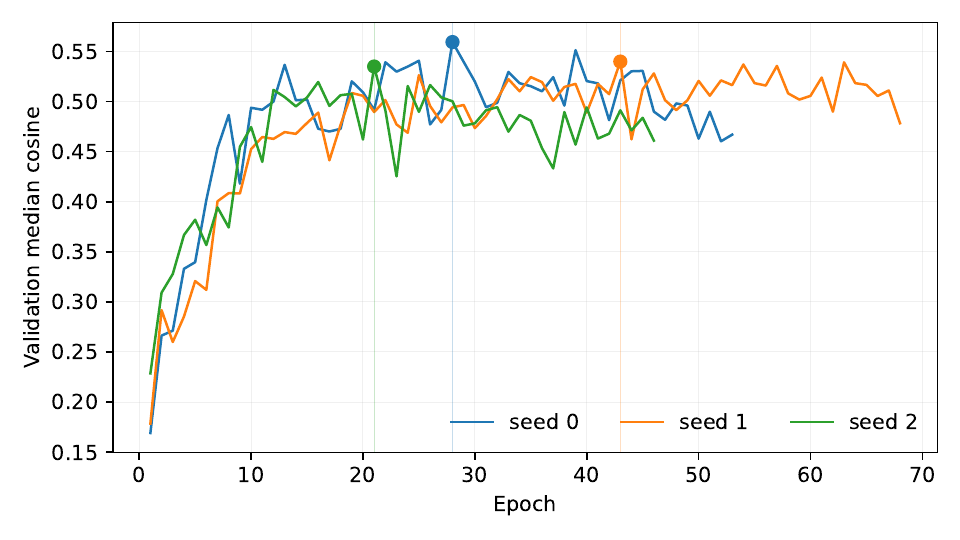}
\caption{Policy-head validation median cosine by epoch. Points mark the selected best epochs (28, 43, and 21 for seeds 0, 1, and 2); training stops after 25 further epochs without improvement (epochs 53, 68, and 46), well below the 150-epoch cap. The rerun exactly reproduces the original best epochs and test medians.}
\label{fig:policy_validation_curves}
\end{figure}

\begin{table}[h]
\centering
\scriptsize
\setlength{\tabcolsep}{3pt}
\renewcommand{\arraystretch}{0.94}
\begin{minipage}[t]{0.49\linewidth}
\centering
\caption{Frozen-readout summary. Intervals are pair-level bootstrap 95\% CIs over 43 test pairs.}
\label{tab:ridge_si}
\begin{tabular}{@{}p{0.56\linewidth}r@{}}
\toprule
Metric & Value \\
\midrule
Selected alpha & 100 \\
Fit / validation / test MAE & 1.623 / 2.214 / 2.259 \\
Existing-head test MAE & 2.087 \\
Original-$h$ fit cosine median & $4.089\times10^{-5}$ \\
Original-$h$ test cosine median & $1.182\times10^{-5}$ \\
$z$-scored fit cosine median & 0.005436 \\
$z$-scored test cosine median & 0.002171 \\
$z$-scored test cosine 95\% CI & $[-0.000467,0.005206]$ \\
Existing-head original-$h$ test cosine median & 0.1074 \\
Two-coordinate original-$h$ weight-energy fraction & 0.999878 \\
\bottomrule
\end{tabular}
\end{minipage}\hfill
\begin{minipage}[t]{0.49\linewidth}
\centering
\caption{PCA summary, fit only on 2{,}855 rows; direction statistics use 43 test pairs.}
\label{tab:pca_si}
\begin{tabular}{@{}p{0.56\linewidth}r@{}}
\toprule
Metric & Value \\
\midrule
$k_{90}$ & 6 \\
Variance captured at $k_{90}$ & 0.901907 \\
First-PC explained variance & 0.623659 \\
Test top-90\% norm-fraction median & 0.818403 \\
Test top-90\% norm-fraction 95\% CI & $[0.771821,0.848506]$ \\
Test top-90\% energy-fraction median & 0.669783 \\
Mean fit direction retained in top-90\% subspace & 0.949503 \\
Test full-space cosine median & 0.284276 \\
Test full-space cosine 95\% CI & $[-0.045306,0.426782]$ \\
Test top-90\% cosine median & 0.281570 \\
Test top-90\% cosine 95\% CI & $[-0.013692,0.440007]$ \\
\bottomrule
\end{tabular}
\end{minipage}
\end{table}

\input{tables/tableS3_policy_seeds.tex}

\input{tables/tableS4_amplitude_supervision.tex}

\input{tables/tableS5_step_sensitivity.tex}

\input{tables/tableS6_policy_decode.tex}

\input{tables/tableS7_policy_decode_sample.tex}

\paragraph{Legacy artifacts retained as text only.}
\label{sec:legacy_artifacts}
Two early visualizations are not reproduced as figures because their embedded annotations conflict with the prespecified statistics. The interpolation figure contained ``65\% mean gap recovered,'' a legacy normalized summary rather than G1's prespecified monotonicity statistic; the rollout figure contained ``mean $\lvert E\rvert$ $2.00\rightarrow0.22$,'' whereas the current 30-goal archive has initial and final median absolute errors $2.000$ and $0.251$ and final mean $0.288$. The current claims use 38/38 monotonicity, median final jump-to-absolute-change ratio $0.806$, and the 30-goal archive.
\begin{figure}[h]
\centering
\includegraphics[width=0.9\textwidth]{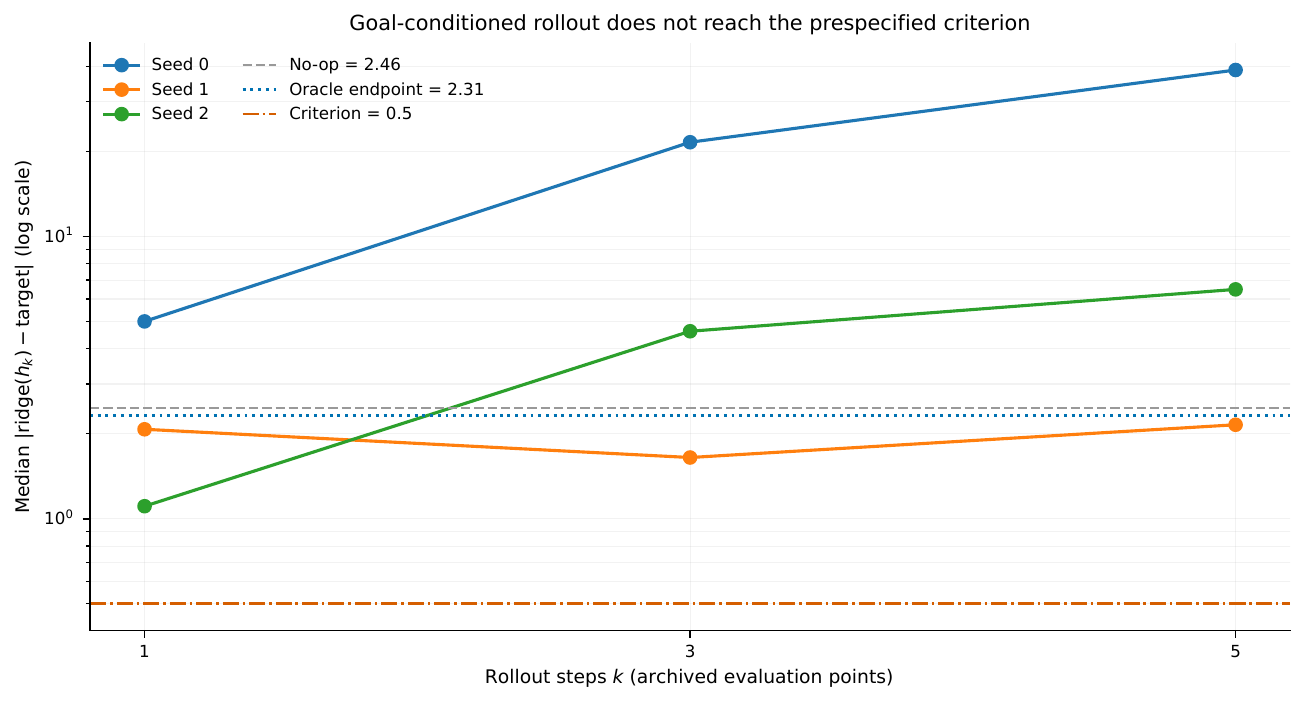}
\caption{Archived goal-conditioned rollout at the saved evaluation points $k=1,3,5$; no values are imputed for $k=2,4$. At $k=5$, seeds 0--2 have median absolute readout errors $38.82$, $2.15$, and $6.49$, respectively, and none reaches the prespecified $<0.5$ criterion. The no-op ($2.46$) and oracle-endpoint ($2.31$) lines expose the readout-reference scale, while seed 0 diverges. This is a checkpoint-specific negative result, not evidence that all goal-conditioned rollout must fail.}
\label{fig:goal_rollout_seeds}
\end{figure}

\begin{figure}[h]
\centering
\includegraphics[width=0.95\textwidth]{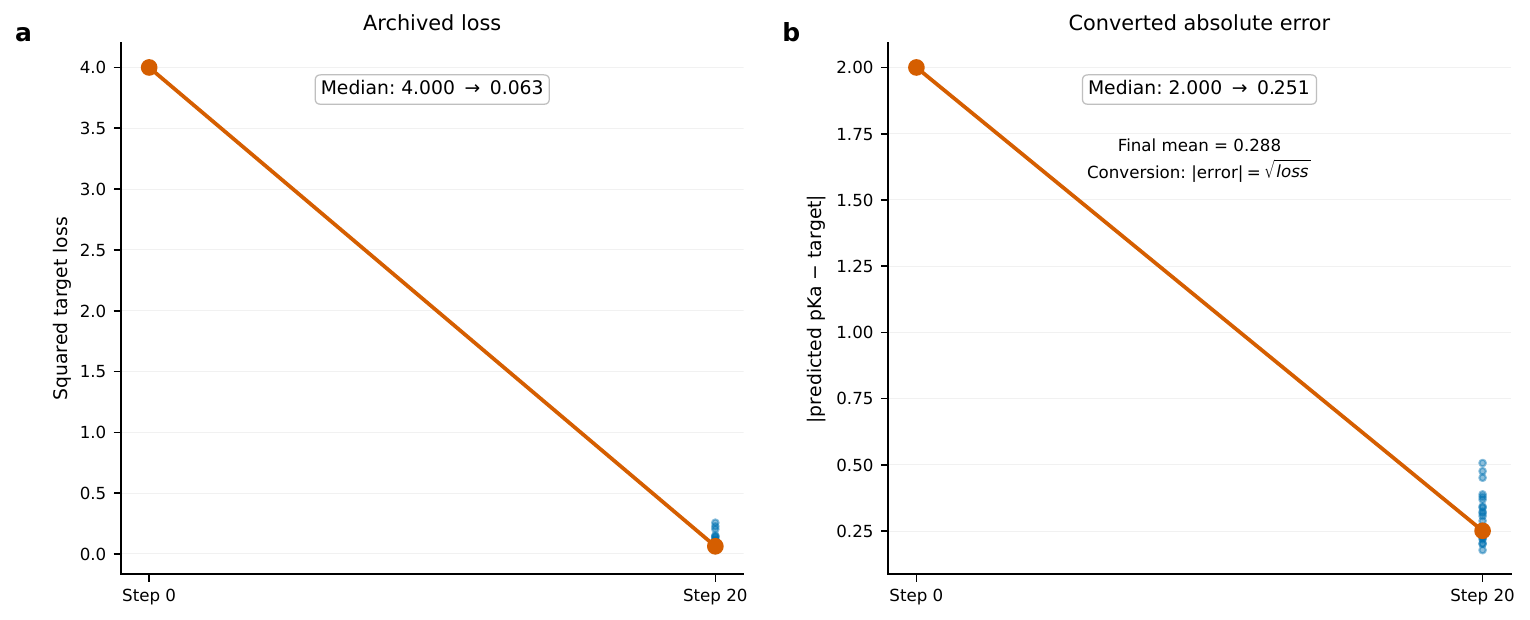}
\caption{G2 squared-loss to absolute-error conversion for 30 archived trajectories. (a) Median squared target loss decreases from $4.000$ to $0.0631$. (b) Applying $|\mathrm{error}|=\sqrt{\mathrm{loss}}$ gives median absolute error $2.000\rightarrow0.251$ and final mean absolute error $0.288$. The archived $0.0631$ is a loss, not an absolute error.}
\label{fig:g2_loss_conversion}
\end{figure}

\begin{figure}[h]
\centering
\includegraphics[width=\textwidth]{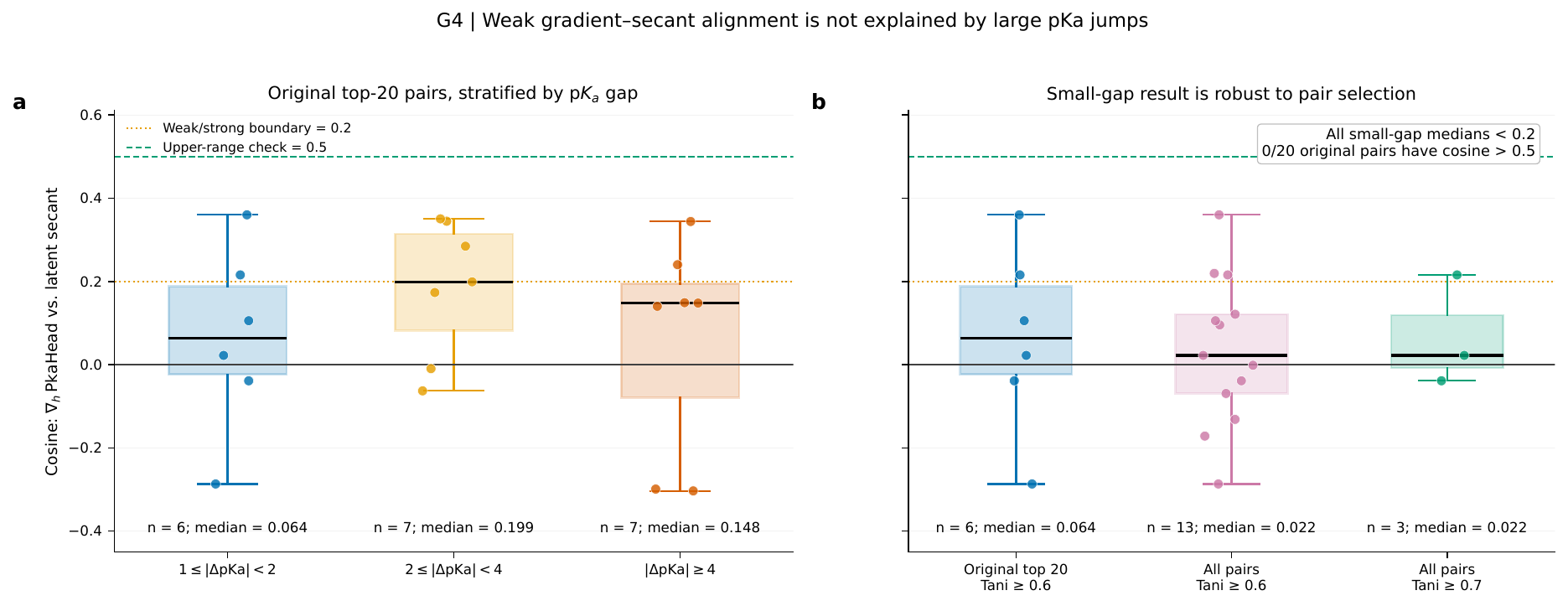}
\caption{Supporting G4 gradient--secant diagnostic. (a) The original top-20 pairs stratified by $|\Delta\pka|$; the small-gap median $0.064$ uses $n=6$. (b) Small-gap results under alternative pair-selection rules; the expanded all-pairs median $0.022$ uses $n=13$. These are different pair banks, not successive estimates on one cohort, and are not directly comparable. Bands are $|\cos|<0.05$ (random), $0.05$--$0.2$ (weak direction), and $\geq0.2$ (strong direction). No original top-20 pair exceeds cosine $0.5$.}
\label{fig:g4}
\end{figure}

\FloatBarrier
\section{Generation audit details}

The SFT training pool contains 2{,}295 rows and no rows are dropped. A direction-correct candidate exists for 81.35\% of rows; the remaining 428 rows use the minimum-target-error candidate. The selected training candidates have direction accuracy $0.8135$, target MAE $1.2825$, and copy rate $0.0157$, with a mean of 7.480 valid candidates per row. These training-pool statistics describe oracle selection and must not be confused with single-sample deployment.

On all 255 evaluation rows, the no-op reference has target MAE $2.2932$. Restricting no-op to the gen2-scored and SFT-scored subsets gives $2.2926$ and $2.2970$, respectively; the main table uses the full-set value. The source-anchor MAE changes from $1.9616$ to $2.1439$, an increase of $0.1822$ that remains below the prespecified $0.3$ limit. Gen2-final Best-of-8 attains direction accuracy $0.7529$ and MAE $1.5000$, compared with $0.8000$ and $1.3128$ after SFT. These are oracle-at-8 values.

\subsection{Offline DPO construction, optimization, and criteria}
\label{sec:dpo_audit}

The four gen2-final training banks contain 2{,}295 rows. Pair construction removes 35 rows with fewer than two unique valid candidates, 37 whose minimum-error chosen structure copies the source, and two whose rejected-minus-chosen absolute-error margin does not exceed $10^{-8}$, leaving 2{,}221 pairs. Chosen and rejected direction accuracies within these training pairs are $0.8154$ and $0.5200$; their copy rates are $0$ and $34/2{,}221=0.0153$. These are preference-data diagnostics, not held-out generator performance. The retained pairs have zero exact-task, ordered-molecule-pair, and source-molecule overlap with the 255 evaluation rows.

The policy and frozen reference start from the same rej1 adapter: all 504 reference LoRA tensors are explicitly cloned, with maximum initial absolute difference $0$. Only the 43{,}646{,}976 policy LoRA parameters train. \texttt{PkaHead} is not instantiated in the DPO model, and the molecular projector remains frozen and is copied byte-for-byte. Training uses $\beta=0.1$, learning rate $10^{-5}$ with cosine decay and 10 warmup steps, and seed 42. The primary run uses two epochs (1{,}112 optimizer steps); the fixed budget ablation uses four epochs (2{,}224 steps) with the same data, initialization, and optimization settings. Four GPUs each process one chosen/rejected pair per step without gradient accumulation, giving four pairs per optimizer step. The final checkpoint is evaluated; no post-result hyperparameter tuning is performed.

The rej1 baseline recomputed from its raw bank has $173/255=0.6784$ single-sample direction accuracy, $13/255=0.0510$ copy rate, and no missing predictions. Both rej1 and DPO use \texttt{do\_sample=True}, temperature $0.9$, top-$p$ $0.95$, at most 96 new tokens, and seed 0; single-sample metrics take the first candidate. DPO has nine missing predictions at two epochs and 29 at four epochs, all scored as source/no-op. The supplementary direct comparison with rej1 has 14 rej1-only and 23 DPO-only correct rows at two epochs (exact McNemar $p=0.1877$) and 29 rej1-only and 18 DPO-only rows at four epochs ($p=0.1439$); these comparisons are supplementary to the primary train-selected constant inference. Two-epoch DPO Best-of-8 attains $195/246=0.7927$ among scored rows and $195/255=0.7647$ when the nine missing rows are no-op; four-epoch DPO Best-of-8 attains $173/226=0.7655$ and $173/255=0.6784$, respectively, both under minimum-target-error oracle selection. A post-hoc held-out-oracle constant sensitivity ($3.28$, direction accuracy $0.6784$) gives $p=0.3742$ at two epochs and, with 54 constant-only and 43 DPO-only rows, $p=0.3099$ at four epochs; it is not the primary test because that constant is selected on the evaluation rows.

\begin{table}[!h]
\centering
\caption{Offline-DPO outcomes on all 255 evaluation rows at the primary two-epoch budget and the fixed four-epoch ablation. The primary inferential test is the train-selected constant comparison; direction and copy values are design references. Best-of-8 is excluded from deployment metrics.}
\label{tab:dpo_criteria}
\footnotesize
\setlength{\tabcolsep}{4pt}
\begin{tabular}{@{}>{\raggedright\arraybackslash}p{0.24\linewidth}>{\raggedright\arraybackslash}p{0.40\linewidth}>{\raggedright\arraybackslash}p{0.16\linewidth}c@{}}
\toprule
Metric & Observed result & Reference & Reading \\
\midrule
Single-sample direction accuracy & $182/255=0.7137$ & $0.72$ design reference & Below reference \\
Train-selected constant comparison & 28 constant-only, 39 DPO-only; exact $p=0.2215$ & $p<0.05$ & Not detected \\
Copy rate & $1/255=0.0039$ & $0.15$ design bound & Within bound \\
\midrule
Four-epoch direction accuracy & $162/255=0.6353$ & $0.72$ design reference & Below reference \\
Four-epoch constant comparison & 45 constant-only, 36 DPO-only; exact $p=0.3742$ & $p<0.05$ & Not detected \\
Four-epoch copy rate & $1/255=0.0039$ & $0.15$ design bound & Within bound \\
\bottomrule
\end{tabular}
\end{table}

\end{document}

%% file: tables/table_training_config.tex
\begin{table}[t]
\centering
\caption{Training configurations and fixed budget ablations. For rejection SFT and DPO, the two epoch counts are the original run and the fixed training-budget rerun; neither uses held-out checkpoint selection. Budget rows retain all 255 rows with missing predictions as source/no-op and use the train-selected constant 3.74.}
\label{tab:training_config}
\small
\setlength{\tabcolsep}{4pt}
\begin{tabular}{@{}p{3.0cm}p{2.6cm}p{1.4cm}p{3.6cm}p{3.3cm}@{}}
\toprule
Component & Initialization & Epochs & Trainable parameters & Selection \\
\midrule
C / B$'$ / B$''$ & LatentChem Stage-2 & 3 & LoRA + \texttt{PkaHead} & Final \\
Rejection SFT & gen2-final & 1 / 3 & LoRA + \texttt{PkaHead} & Final \\
Offline DPO & Rejection SFT (1 epoch) & 2 / 4 & 43.6M LoRA & Final \\
Policy head & Scratch & $\le 150$; patience 25 & MLP & Validation median cosine \\
\bottomrule
\end{tabular}

\vspace{0.5em}
\begin{tabular}{@{}llrrrr@{}}
\toprule
Model & Epochs & Scored & Direction & $p$ vs constant & Copy rate \\
\midrule
Rejection SFT & 1 & 252/255 & 0.6863 & 0.7239 & 0.1216 \\
Rejection SFT & 3 & 242/255 & 0.6510 & 0.6488 & 0.0784 \\
\midrule
Offline DPO & 2 & 246/255 & 0.7137 & 0.2215 & 0.0039 \\
Offline DPO & 4 & 226/255 & 0.6353 & 0.3742 & 0.0039 \\
\bottomrule
\end{tabular}
\end{table}

%% file: tables/table5_direction_amplitude.tex
\begin{table}[t]
\centering
\caption{Directional learnability and amplitude controllability evaluated on the same 43 held-out pairs. The direction criterion applies to the median point estimate ($>0.3$), not to every pair or to the CI lower bound; pair-level fractions and median CIs are given in Table~S3. The amplitude criterion is signed correlation $>0.5$. Amplitude correlations are invariant to the fixed step size. G4 and ridge have no comparable target-conditioned amplitude evaluation.}
\label{tab:direction_amplitude}
\small
\begin{tabular}{lrrcc}
\toprule
Method & Direction cos & Amplitude corr & Direction & Amplitude \\
\midrule
G4 naive gradient & 0.107 & --- & \xmark & --- \\
Frozen ridge ($z$-scored) & 0.00217 & --- & \xmark & --- \\
Policy seed 0 & 0.482 & -0.079 & \cmark & \xmark \\
Policy seed 1 & 0.539 & 0.208 & \cmark & \xmark \\
Policy seed 2 & 0.469 & -0.012 & \cmark & \xmark \\
Policy ensemble & 0.577 & -0.020 & \cmark & \xmark \\
\bottomrule
\end{tabular}
\end{table}

%% file: tables/table4_generation.tex
\begin{table*}[t]
\centering
\caption{Generation results with separated evaluation protocols. The all-row direction comparison uses the same 255 rows for gen2-final, SFT, and the train-selected constant $3.74$, with missing single-sample predictions treated as no-op; SFT does not have a detected advantage over the constant (34 constant-only versus 38 SFT-only correct, exact McNemar $p=0.7239$). The separate paired gen2-final/SFT comparison uses the 235 both-scored rows (direction $p=0.00418$); its MAE cells therefore share one denominator. SFT Best-of-8 is an oracle diagnostic on all 255 rows, not a deployable selector. The strict no-op reference has no direction prediction and is N/A rather than a direction baseline.}
\label{tab:generation}
\small
\begin{tabular}{@{}p{0.29\linewidth}cccc@{}}
\toprule
Metric & gen2-final & SFT single & SFT Best-of-8 & Train-selected constant \\
\midrule
Direction accuracy (all $n=255$; missing=no-op) & 0.5922 & 0.6863 & 0.8000 (oracle) & 0.6706 \\
Direction accuracy (paired $n=235$) & 0.6426 & \textbf{0.6936} & --- & --- \\
Target MAE (paired $n=235$) & 2.424 & \textbf{2.190} & --- & --- \\
Best-of-8 target MAE (oracle $n=255$) & --- & --- & 1.313 & --- \\
Unique valid molecules & 227 & 241 & --- & --- \\
Copy rate & 0.051 & 0.122 & --- & --- \\
\midrule
\multicolumn{5}{l}{Separate no-op reference (n=255): strict direction N/A; target MAE = 2.293.} \\
\bottomrule
\end{tabular}
\end{table*}

%% file: tables/tableS0_text_shortcut.tex
\begin{table}[h]
\centering
\caption{Pre-repair text-shortcut diagnosis. The C PkaHead value $1.103$ is recomputed from 24 held-out diagnostic rows; repaired C is evaluated on a different cohort of 1{,}151 test rows after two combined repairs. Modality-conflict values are archived aggregate MAEs ($n=24$), with source and donor ground truth (GT) scored separately. Per-row conflict predictions were not preserved, so no intervals or output-invariance counts are reported. The pre/post values are not a controlled ablation.}
\label{tab:shortcut}
\scriptsize
\setlength{\tabcolsep}{4pt}
\begin{tabular}{@{}>{\raggedright\arraybackslash}p{0.28\linewidth}>{\raggedright\arraybackslash}p{0.12\linewidth}>{\raggedright\arraybackslash}p{0.24\linewidth}rr@{}}
\toprule
Condition & Query SMILES & Cohort / evidence & Source MAE & Donor MAE \\
\midrule
C pre-repair, \texttt{PkaHead} & present & 24-row probe; recomputed & 1.103 & --- \\
C repaired, combined repairs & removed & 1{,}151-row test; recomputed & 2.087 & --- \\
\midrule
Conflict: base & source & 24-row archived aggregate & 1.17 & 4.06 \\
Conflict: text $\rightarrow$ donor; molecule $\rightarrow$ source & donor & 24-row archived aggregate & 4.08 & 1.20 \\
Conflict: features $\rightarrow$ donor; text $\rightarrow$ source & source & 24-row archived aggregate & 1.17 & 4.07 \\
\bottomrule
\end{tabular}
\end{table}

%% file: tables/tableS3_policy_seeds.tex
\begin{table}[h]
\centering
\caption{Target-conditioned policy-head seed detail on 43 held-out pairs. The prespecified direction criterion applies to the median point estimate, not to every pair or to the CI lower bound; all four median CIs have lower bounds below $0.3$. Small-gap pairs are the 13 pairs with $1\leq|\Delta\pka|<2$. Amplitude statistics use $s=1.0$; correlation and sign accuracy do not depend on the scale of a fixed nonzero step. Norm CV is the coefficient of variation of the raw policy displacement norms before fixed-step normalization.}
\label{tab:policy_seeds}
\scriptsize
\begin{tabular}{lccccccc}
\toprule
Seed & Cos & 95\% CI & Pairs $\cos>0.3$ & Small-gap cos & Amp. corr & Sign acc. & Norm CV \\
\midrule
Seed 0 & 0.482 & [0.292, 0.725] & 27/43 (0.628) & 0.463 & -0.079 & 0.395 & 0.722 \\
Seed 1 & 0.539 & [0.274, 0.691] & 26/43 (0.605) & 0.496 & 0.208 & 0.744 & 0.733 \\
Seed 2 & 0.469 & [0.287, 0.673] & 27/43 (0.628) & 0.344 & -0.012 & 0.512 & 0.638 \\
Ensemble & 0.577 & [0.235, 0.695] & 27/43 (0.628) & 0.442 & -0.020 & 0.395 & 0.710 \\
\bottomrule
\end{tabular}
\end{table}

%% file: tables/tableS4_amplitude_supervision.tex
\begin{table}[h]
\centering
\caption{Explicit amplitude supervision with $\lambda_{\mathrm{amp}}=1$. The joint criterion requires median cosine $\geq0.3$ and learned-amplitude correlation $>0.5$. Neither any individual seed nor the ensemble satisfies both criteria. These values use the learned per-pair displacement norm, unlike the fixed-step sensitivity analysis.}
\label{tab:amplitude_supervision}
\scriptsize
\begin{tabular}{lccccc}
\toprule
Seed & Cos & Learned amp. corr & Median $|\Delta$pKa$|$ & Median step norm & Joint pass \\
\midrule
Seed 0 & 0.312 & -0.320 & 2.619 & 1.784 & \xmark \\
Seed 1 & 0.012 & 0.281 & 11.541 & 3.662 & \xmark \\
Seed 2 & 0.232 & 0.809 & 2.787 & 1.630 & \xmark \\
Ensemble & 0.192 & 0.126 & 5.535 & 1.492 & \xmark \\
\bottomrule
\end{tabular}
\end{table}

%% file: tables/tableS5_step_sensitivity.tex
\begin{table}[h]
\centering
\caption{Fixed-step amplitude sensitivity for the unit-normalized policy ensemble on 43 held-out pairs. The target band is $0.2\leq|\Delta\pka|\leq1.0$. Signed Pearson correlation and sign accuracy are invariant to positive fixed rescaling. The calibration condition requires both a median in the target band and correlation $>0.3$; it is not met at any step.}
\label{tab:step_sensitivity}
\small
\begin{tabular}{rcccc}
\toprule
Step $s$ & Median $|\Delta\pka|$ & Fraction in $[0.2,1.0]$ & Sign acc. & Signed corr \\
\midrule
0.1 & 0.0296 & 0.023 & 0.395 & -0.0199 \\
0.2 & 0.0592 & 0.116 & 0.395 & -0.0199 \\
0.3 & 0.0888 & 0.279 & 0.395 & -0.0199 \\
0.5 & 0.1481 & 0.442 & 0.395 & -0.0199 \\
1.0 & 0.2961 & 0.419 & 0.395 & -0.0199 \\
2.0 & 0.5922 & 0.419 & 0.395 & -0.0199 \\
3.0 & 0.8883 & 0.419 & 0.395 & -0.0199 \\
\bottomrule
\end{tabular}
\end{table}

%% file: tables/tableS6_policy_decode.tex
\begin{table}[h]
\centering
\caption{Greedy P3 policy-direction decoding on all 80 eligible held-out source molecules. Paired effects are intervened minus baseline learned-oracle predictions, so unchanged sequences have zero effect and count against directional success. Identical outputs retain that zero effect by identity when their shared invalid SMILES cannot be oracle-scored; no changed output lacks a paired oracle score. Checks are validity $\geq0.9$, Tanimoto median $\geq0.6$, direction $\geq0.65$, and median $|\Delta\pka|\in[0.2,1.0]$. The MAE-1.1 PkaHead is a learned oracle, not external or experimental validation. Property/generator PkaHead and projector maximum absolute tensor differences are 0.2456 and 0.0574, respectively.}
\label{tab:p3_decode}
\small
\begin{tabular}{lrc}
\toprule
Metric & Value & Check \\
\midrule
$h_T$ target cosine, median & 1.0000 & --- \\
$h_T$ relative error, median & 0.0081 & --- \\
Generator-readout direction & 0.062 & --- \\
Generator-readout median $|\Delta\pka|$ & 0.0000 & --- \\
Intervened SMILES validity & 1.000 & \cmark \\
Source-anchored Tanimoto, median & 0.576 & \xmark \\
Changed decoded sequences & 1/80 & --- \\
Both learned-oracle scored & 80/80 & --- \\
Paired learned-oracle direction & 0.000 & \xmark \\
Paired learned-oracle median $|\Delta\pka|$ & 0.0000 & \xmark \\
Paired learned-oracle fraction in $[0.2,1.0]$ & 0.000 & --- \\
\bottomrule
\end{tabular}
\end{table}

%% file: tables/tableS7_policy_decode_sample.tex
\begin{table}[h]
\centering
\caption{Sampled ($T=0.9$, top-$p=0.95$, seed 0) P3 policy-direction decoding on all 80 eligible held-out source molecules. Paired effects are intervened minus baseline learned-oracle predictions, so unchanged sequences have zero effect and count against directional success. Identical outputs retain that zero effect by identity when their shared invalid SMILES cannot be oracle-scored; no changed output lacks a paired oracle score. Checks are validity $\geq0.9$, Tanimoto median $\geq0.6$, direction $\geq0.65$, and median $|\Delta\pka|\in[0.2,1.0]$. The MAE-1.1 PkaHead is a learned oracle, not external or experimental validation. Property/generator PkaHead and projector maximum absolute tensor differences are 0.2456 and 0.0574, respectively.}
\label{tab:p3_decode_sample}
\small
\begin{tabular}{lrc}
\toprule
Metric & Value & Check \\
\midrule
$h_T$ target cosine, median & 1.0000 & --- \\
$h_T$ relative error, median & 0.0081 & --- \\
Generator-readout direction & 0.062 & --- \\
Generator-readout median $|\Delta\pka|$ & 0.0000 & --- \\
Intervened SMILES validity & 0.938 & \cmark \\
Source-anchored Tanimoto, median & 0.514 & \xmark \\
Changed decoded sequences & 0/80 & --- \\
Both learned-oracle scored & 75/80 & --- \\
Paired learned-oracle direction & 0.000 & \xmark \\
Paired learned-oracle median $|\Delta\pka|$ & 0.0000 & \xmark \\
Paired learned-oracle fraction in $[0.2,1.0]$ & 0.000 & --- \\
\bottomrule
\end{tabular}
\end{table}